\documentclass[12pt]{article}

\def\be{\begin{equation}}
\def\ee{\end{equation}}

\usepackage{amssymb}
\usepackage{amsmath}
\usepackage{epsfig}
\usepackage{psfrag}

\begin{document}
\begin{titlepage}

\thispagestyle{empty}
\hskip 1 cm
\vskip 0.5cm

\vspace{25pt}
\begin{center}
    { \LARGE{\bf Volume Weighted Measures of Eternal Inflation in the
    Bousso-Polchinski Landscape}}
    \vspace{33pt}

  {\large  {\bf   Timothy Clifton\footnote{T.Clifton@cantab.net},
  Stephen Shenker\footnote{SShenker@stanford.edu}
  and Navin Sivanandam\footnote{NavinS@stanford.edu}}}

    \vspace{15pt}

 {Department of Physics,
    Stanford University, CA 94305, USA}

\date{{\normalsize {\today}}}

  \end{center}

\begin{abstract}

We consider the cosmological dynamics associated with volume weighted
measures of eternal inflation, in the Bousso-Polchinski model of the string theory
landscape.  We find that this measure predicts that observers are most
likely to find themselves in low energy vacua with one flux
considerably larger than the rest.  Furthermore, it allows for a
satisfactory anthropic explanation of the cosmological constant problem by
producing a smooth, and approximately constant, distribution of
potentially observable values of $\Lambda$.  The low energy vacua
selected by this measure are often short lived.  If we require anthropically acceptable vacua to
have a  minimum life-time of 10 billion years, then for reasonable
parameters a typical observer should expect their vacuum to have a
life-time of approximately 12 billion years.  This prediction is model dependent,
but may point toward a solution to the coincidence problem of cosmology.

\end{abstract}

\vspace{10pt}
\end{titlepage}

\tableofcontents

\newpage

\section{Introduction}

There is now strong evidence that string theory contains a large
number of metastable vacua, collectively referred to as the
landscape \cite{land}.  In order to make progress in understanding the place of a typical
observer in this landscape of possibilities we must understand which vacua are possible, and the method of
their population.  The former of these problems is a current topic of
research  and we will not attempt to contribute to this field here.
Instead we will take the pioneering model of Bousso
and Polchinski \cite{Bousso:2000xa}, and investigate the latter.  

The population mechanism for the landscape is eternal inflation.  De Sitter vacua are constantly
expanding and, in string theory, are meta-stable.  As these vacua expand they nucleate
bubbles of new vacuum, which themselves expand and nucleate further
bubbles.  In this way it is expected that all corners of
the landscape are realised\footnote{For a recent study of
  possible mechanisms to divide the landscape into different regions
  see \cite{islands}.}, and the pertinent question is
then which are the most probable.  The fundamentals involved in
answering this question are currently a matter of some contention:  A
difficulty often referred to as `the measure problem.'

Various different measures have been proposed in the literature.  We
will not attempt to describe them all here, nor compare their various
implicit benefits and difficulties.  Instead, we will take one particular class
of measures, and investigate the outcome of using them to predict the
likely vacua that observers will find themselves in.  The class we
will study are the volume weighted measures, proposed by Linde \cite{volumem}.  This
class embodies the appealing intuition of rewarding the fastest
expanding vacua.  A refinement of this measure has
recently been shown to produce results
with surprisingly little dependence on the choice of time
parameterization, and that do not exhibit a hotness (or youngness) paradox
\cite{andreinew}.   Using the Bousso-Polchinski landscape to
investigate the results of volume-weighted eternal inflation has the
advantage of this model already having been used to investigate
other measures, namely the bubble-counting measure of \cite{bubbles, Schwartz-Perlov:2006hi}
and the holographic measure of \cite{bousso, boussoyang}.  This will allow us to
directly compare the results of several  candidate
measures within the same theoretical framework. 

 The Bousso-Polchinski landscape has
inherent in it a number of approximations.  For example, in this toy model
the charges and their associated moduli are independent of fluxes, which is unlikely to be
true of the real landscape.  The contribution of slow-roll inflation is also ignored.
 Within the Bousso-Polchinski framework we will make several further approximations, for 
 analytic convenience, such as treating discrete fluxes as continuous variables. 
 Nonetheless we will be able to find, within this model,  reliable predictions 
 for the distribution of the cosmological constant, the
likely configuration of fluxes and even the typical lifetime of a
`pocket universe'.

In section 2 we will give a brief account of volume weighted eternal
inflation and, using examples, will give an account of how to
calculate probabilities in this measure.  We will follow this with a
description of the Bousso-Polchinski landscape, and the transition
rates expected within it.  We will then reflect on some anthropic
considerations that can be applied to this landscape.  In particular,
we consider anthropic bounds on the cosmological constant and on
the lifetime of vacua that must be obeyed if observers are to be
expected to occur.

Section 3 gives a description of how the high energy vacua in this
landscape are expected to be populated in volume weighted inflation.
We discuss here how initial conditions are quickly forgotten, once the
highest energy, most rapidly expanding vacua are realised, and
subsequently come to
dominate the evolution of the universe.  We then argue that above a
certain high energy boundary the distribution of probability is
approximately constant, and isotropically distributed.

In section 4 we consider tunneling events down from this high-energy
barrier to a surface one jump up from the anthropic shell.  We find
that the primary contribution to probability on this low energy shell
is due to chains of transitions occurring in a straight line, in one dimension of
the flux space.  Approximating the discrete space as continuous, we
then find an explicit expression for the total tunneling rate down
from high energies to the penultimate sphere, before the anthropic
one.  We find that there exist peaks of probability on this
penultimate sphere, and find their position at various different
levels of approximation.

Section 5 gives a description of the final transition down into the
anthropic shell.  This transition is treated separately as the effects
of gravity are negligible for this final jump.  We find that this
last step also produces peaking, but that this peaking is negligible
in comparison to that found in section 4.

A description of the extent of peaking, due to suppressed transitions,
can be found in section 6.  For a typical set of parameters we find
 that $99.9999\%$ of the vacua in
the anthropic shell should be expected to lie within a fraction $\sim 10^{-23}$
of its surface area.  The selected vacua have one flux much larger than the others.
There are ample vacua in the selected set to allow for an anthropic solution of the cosmological 
constant problem.   We also consider in this section the effect of
the anthropic life-time bound, already discussed in section 2.  The
peak of probability is found to be centered on vacua with
insufficiently long life-times for observers to appear.  If we
consider only vacua with life-times $>10$ billion years, then the
peaking in probability exerts a pressure favoring the vacua with
shorter lives.  We find that a typical vacuum, in which observers take
$10$ billion years to appear, will be most likely to decay
approximately $2.2$ billion years later.

Section 7 gives a discussion of the results obtained, and how they may
be altered by effects that were not included in our model.  For
example, if multiple branes are allowed to be nucleated simultaneously
then the peaking we have found should be expected to be reduced, and
to vanish in the limit of indefinitely many branes nucleating
simultaneously.  We also consider how the required stage of slow-roll
inflation, before re-heating, could modify our results.

\section{The Model}

We consider here volume-weighted measures of inflation in the Bousso-Polchinski
landscape. We will begin with a brief explanation of these measures
in eternal inflation, and then describe the features of the
Bousso-Polchinski landscape that are of relevance for our study.

\subsection{Volume Weighting}

The essential feature of volume weighted measures is that they
reward those regions which expand most rapidly.  That is, in a given
increment of time, $\Delta t$, the 3-volume of each type of vacuum
is increased as a function of its rate of expansion, in some
proportionate way.  Clearly the way in which time is measured is of
primary importance here.  For definiteness we will define `time' as
being the proper time measured by a congruence of comoving
observers.  This can be measured in an invariant way in units of the
11-dimensional Planck time, or by equipping each observer with a
suitably constructed light clock.  Inflation occurs in each vacuum
until a tunneling event creates a bubble of new vacuum in the old
(or on rare occasions until Minkowski space is reached, if this
exists in the theory).  The number of points in any particular
vacuum that do not eventually undergo a transition to a bubble of new
vacuum  are a set of measure zero.  However, due to the
rapid expansion of high-energy vacua, such vanishingly small sets
can still come to dominate the evolution of the universe.  To make
these ideas concrete, we will illustrate this behavior with an example.

\begin{figure}[t]
\center \epsfig{file=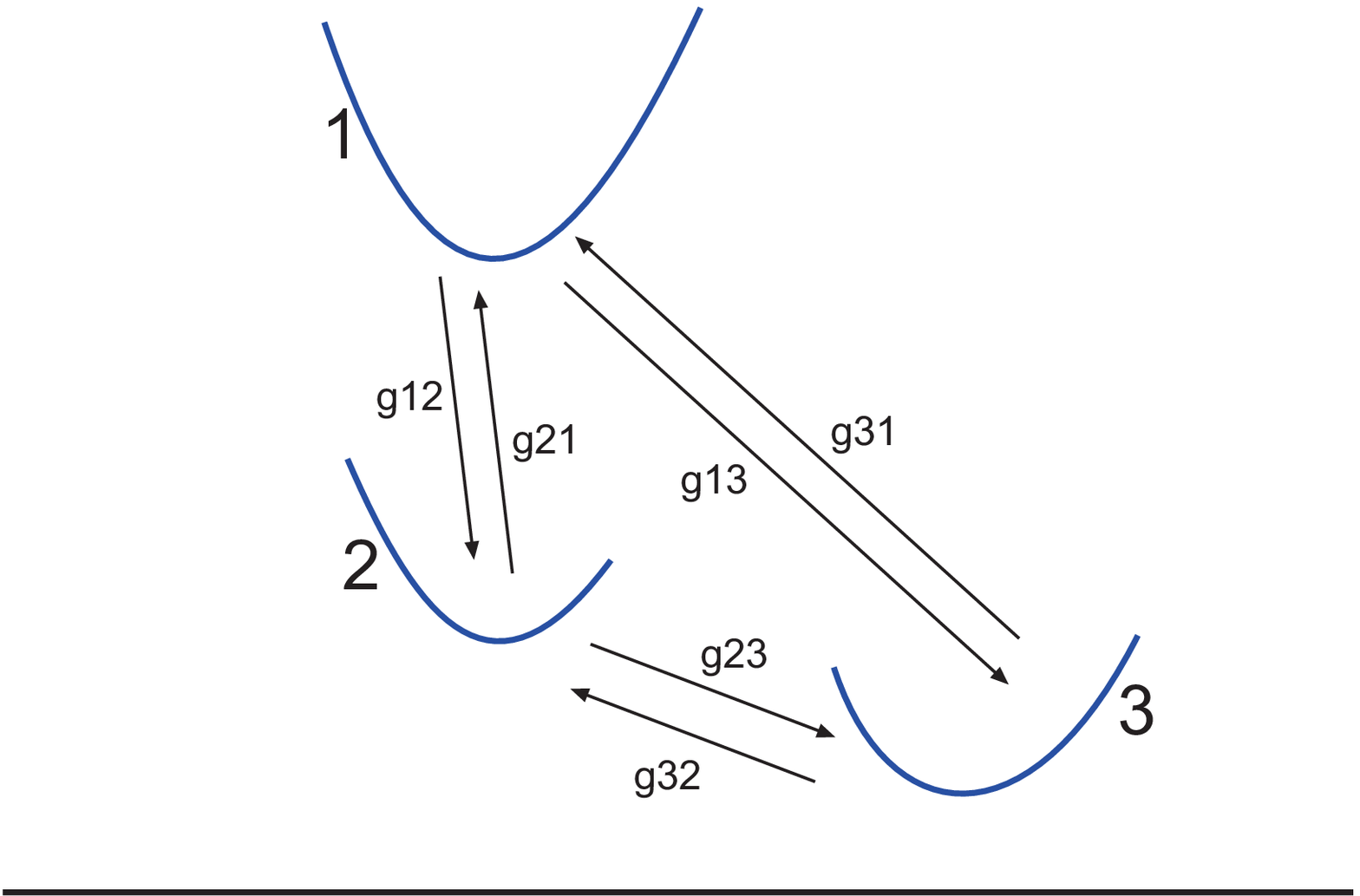,height=8cm} \caption{A toy landscape,
with three dS vacua.  $\kappa_{ij}$ denotes the tunneling rate from
vacuum $i$ to vacuum $j$.} \label{3ds}
\end{figure}

Consider a simple landscape of 3 de Sitter vacua, as shown in Figure
\ref{3ds}.  Here $\kappa_{ij}$ denotes the transition rate from vacuum
$i$ to vacuum $j$, and the $H_i$ are Hubble constants.  We assume that
vacuum 1 has a much higher cosmological constant than vacua 2 and 3,
so that $H_1\gg H_{2,3}$ and $\kappa_{1i}\gg\kappa_{i1},\
\kappa_{23},\ \kappa_{32}$.
\psfrag{1}{$1$} \psfrag{2}{$2$} \psfrag{3}{$3$}
\psfrag{g12}{$\kappa_{12}$} \psfrag{g13}{$\kappa_{13}$}
\psfrag{g21}{$\kappa_{21}$} \psfrag{g31}{$\kappa_{31}$}
\psfrag{g23}{$\kappa_{23}$} \psfrag{g32}{$\kappa_{32}$}
If $P_i$ is the (unnormalized) probability of finding a given point
in the vacuum state $i$, then we can describe this dynamical system
through a set of coupled ODEs \cite{vil}, \cite{Linde:2006nw}:
\begin{eqnarray}
\dot{P_1}&=&\kappa_{21}P_2+\kappa_{31}P_3-(\kappa_{12}+\kappa_{13})P_1+3H_1P_1\nonumber\ ,\\
\dot{P_2}&=&\kappa_{12}P_1+\kappa_{32}P_3-(\kappa_{23}+\kappa_{21}) P_2+3H_2P_2\nonumber\ ,\\
\dot{P_2}&=&\kappa_{13}P_1+\kappa_{23}P_2-(\kappa_{32}+\kappa_{31}) P_3+3H_3P_3\nonumber .
\end{eqnarray}
As long as $H_1\gg\kappa_{ij}$ for all $i$, $j$ (i.e. the highest energy vacuum
expands much faster than it decays), the first of these equations
becomes
\be
\nonumber
\dot{P_1} \simeq 3H_1P_1 \qquad \qquad \Rightarrow \qquad \qquad P_1
\propto e^{-3 H_1 t}.
\ee
The terms dominating the RHS of the second two equations are then
those which are proportional to $P_1$, as $t\rightarrow \infty$.
These equations are solved by
\begin{equation*}
\frac{P_i}{P_1}\simeq \frac{\kappa_{1i}}{3 H_1} \ll 1 \qquad \qquad
\text{or} \qquad \qquad \frac{P_i}{P_j}\simeq \frac{\kappa_{1i}}{\kappa_{1j}}.
\end{equation*}
This confirms that the probability of finding a point in any particular vacuum is
proportional to its rate of decay from the highest energy
(fastest-growing) vacuum.  Essentially, the most rapidly expanding
vacuum sources all of the others -- it is easier to
make the lower vacua by expanding in vacuum 1 and then decaying down,
than it is to do so by simply expanding with the smaller Hubble
constant of the lower vacua.

These results readily generalize to more complicated systems.
First, consider a landscape in which the highest energy vacuum can
decay to all other vacua.  In this case we have
\begin{equation}
\frac{P_i}{P_h}\simeq \frac{\kappa_{hi}}{3 H_h} \ll 1 \qquad \qquad
\text{or} \qquad \qquad \frac{P_i}{P_j}\simeq \frac{\kappa_{hi}}{\kappa_{hj}},
\end{equation}
where $h$ labels the highest energy vacuum.  Now consider a
landscape in which there exist vacua that cannot be created from
direct decays from the highest energy state.  In this case the probability of finding
a point in the vacuum state $j$ is
\begin{equation}
\label{gentun}
\frac{P_j}{P_h}= \sum_{\stackrel{all}{paths}} \frac{\kappa_{ha}}{3 H_h} \times \frac{\kappa_{ab}}{3 H_h}
\times \frac{\kappa_{bc}}{3 H_h} \times ...\times
\frac{\kappa_{ij}}{3 H_h} \ll1.
\end{equation}
The chain of $\kappa_{ij}/3H_h$ gives the probability of starting in the
state $h$ and finishing in $j$, following some series of decays
through the vacua $a$, $b$, $c...$.  To obtain the total probability
of finding a point in the vacuum $j$ it is then necessary to sum over
all possible paths.  Clearly some paths will provide more of a contribution than
others to this sum.  In fact, as long as $\kappa_{ij}\ll H_h$, only the shortest
decay chains will give a significant contribution.  Every chain with
additional, unnecessary, intermediary vacua will be suppressed by
factors of $\kappa_{ij}/3H_h\ll 1$.  However, if $\kappa_{ij} \sim 3 H_h$ then
the products of decay are no longer suppressed.  In this case the
progeny of decaying vacua will be as equally abundant as the vacua
from which they decayed.  Rapid transitions of this type will not
contribute suppressing factors to the decay chains mentioned above.

\subsection{The Bousso-Polchinski Landscape and Transitions}

We will consider here the Bousso-Polchinski (BP) landscape of M-Theory
\cite{Bousso:2000xa}.  This model has the advantages of being
relatively simple and, most importantly for our purposes, having known
expressions for the transition rates.

The BP landscape consists of a discrete set of vacua, each with a
different effective cosmological constant given by
\begin{equation}\label{cc}
\Lambda=-\Lambda_0+\frac{1}{2} \sum_{i=1}^Jq_i^2n_i^2,
\end{equation}
where $J$ is the number of fluxes, $q_i$ the charge associated with each
of them and $n_i$ the number of units of a particular flux. The
discreteness of the $n_i$ arises from the quantization of four-form
fluxes in M-theory compactifications, and their multiplicity from the
geometry of the M-theory compactification -- each non-trivial
three-cycle giving rise to a different four-form flux. The number of
such cycles is generically quite large.  This multiplicity allows a much finer
spacing of $\Lambda$ than that given by a single flux alone, which
typically has a charge only a few orders of magnitude below the Planck scale
\cite{Bousso:2000xa}.

From equation (\ref{cc}) it is clear that for a given choice of
$\Lambda$, the fluxes are constrained to lie on a sphere in
flux space, of radius $\sqrt{2 (\Lambda+\Lambda_0)}$.
Taking $0\leqslant \Lambda \lesssim 10^{-120}$ (in 4D Planck
units) we obtain a thin shell of radius $\sim \sqrt{2 \Lambda_0}$, which
corresponds to the set of vacua in the BP landscape which have an
anthropically acceptable cosmological constant (see below).

Transitions in this landscape occur via the nucleation of branes. The
transition rates, including gravitational effects, are given by
\cite{Coleman:1980aw}
\begin{equation}
\kappa_{ij}=A_{ij}e^{-B_{ij}}\ .
\end{equation}
The $A_{ij}$ can be approximated by $A_{ij}\sim1$ (any effect they
have will be exponentially unimportant). Now let us consider the
nucleation of bubbles which change the flux through some particular
cycle by one unit, from $n_i$ to $n_i-1$, such that
\begin{equation}
\nonumber
|\Delta\Lambda_i|= (n_i-1/2)q_i^2\ .
\end{equation}
The tunneling rate is then given as the product of its value in
flat space, times some gravitational correction, as \cite{Park}
\begin{equation}
\label{tunnel}
B_{i\downarrow}=B^{\textrm{flatspace}}_{i\downarrow}r(x,y),
\end{equation}
where the gravitational correction factor is
\be
r(x,y) = \frac{2[(1+xy)-(1+2xy+x^2)^{1/2}]}{x^2(y^2-1)(1+2xy+x^2)^{1/2}},
\ee
and the flat space rate is \cite{brown}
\be
B^{\textrm{flat-space}}_{i\downarrow}= \frac{27\pi^2}{8}\frac{1}{(n_i-1/2)^3q_i^2}.
\ee
The dimensionless parameters in $r(x,y)$ are given by
\be
x\equiv\frac{3q_i^2}{8|\Delta\Lambda_i|}=\frac{3}{8(n_i-1/2)}
\qquad \qquad
\text{and}
\qquad \qquad
y\equiv\frac{2\Lambda}{|\Delta\Lambda_i|}-1\ .
\ee
For our purposes $n_i$ will always be $\gg 1$, so that $x \simeq 3/8n
\ll 1$.  The variable $y$ has a clear interpretation as the
energy-level of the parent vacuum.  At high energies $\Lambda_i \gg
|\Delta\Lambda_i|$, so that $y \gg 1$.  Conversely, at the lowest
energies $\Lambda_i \sim |\Delta\Lambda_i|$, so that $y \sim 1$.  The
combination $x y$ can be interpreted as a measure of distance from the
axes of the flux space.  Consider the surface defined by $x y =3/8$, we then have
\be
\nonumber
xy \simeq \frac{3}{8 n_i} \frac{2\Lambda_i}{|\Delta\Lambda_{ij}|} \simeq
\frac{3}{4} \frac{\Lambda}{n_i^2 q_i^2} \simeq \frac{3}{8} \qquad
\Rightarrow \qquad \sum_{k \neq i} n_k^2 q_k^2 \simeq 2 \Lambda_0.
\ee
The LHS of this second equation is the perpendicular distance
(squared) away from the axis corresponding to the direction in which
the transition is occurring.  The RHS is the radius (squared) of the
anthropic shell.  Therefore when $x y \lesssim 3/8$ the transition is
occurring in a cylinder around this axis, of the same cross-sectional
area as the anthropic shell.  When $x y \gg 3/8$ the transition is
occurring well away from the axis.

Useful estimates of tunneling rates can be found in a number of
different regimes \cite{Schwartz-Perlov:2006hi}.  Firstly, at high
energies far from the axis, when $y \gg 1$ and $xy \gg 1$, it can be seen that
\be
\label{r1}
\qquad r(x,y) \simeq \sqrt{2} (xy)^{-3/2} \ll1.
\ee
The tunneling rate (\ref{tunnel}) is then given by
\be
\label{high1}
\kappa_{i j} \simeq \textrm{Exp}\left[- \frac{3 \sqrt{6} \pi^2
    q_i}{\Lambda_i^{3/2}}\right].
\ee
This expression is independent of $n_i$.
Alternatively, closer to the axis, when $y \gg 1$ and $xy \ll 1$, we
    obtain the Pad\'{e} approximant
\be
\label{r2}
r(x,y) \simeq \frac{1-x y}{1+x y}.
\ee
This approaches the flat space limit as $x y\rightarrow 0$:
\begin{equation}
\label{gamlast}
\kappa_{ij} \sim \textrm{Exp}\left[-\frac{27\pi^2}{8}\frac{1}{n_i^3q_i^2} \right].
\end{equation}

Another  regime of interest is at low energy, when $y \sim 1$: that of
the final transition, down into the anthropic shell.  In this limit \cite{Coleman:1980aw}
\begin{equation}
\nonumber
r(x,y) \simeq (1+x^2)^{-2} = \frac{(n_i-1/2)^2}{(n_i-1/8)^2} \simeq 1,
\end{equation}
such that the gravitational correction is vanishingly small.  As long as this
condition is met, we can then approximate the tunneling rate down into
the anthropic shell by equation (\ref{gamlast}).

\begin{figure}[t]
\center \epsfig{file=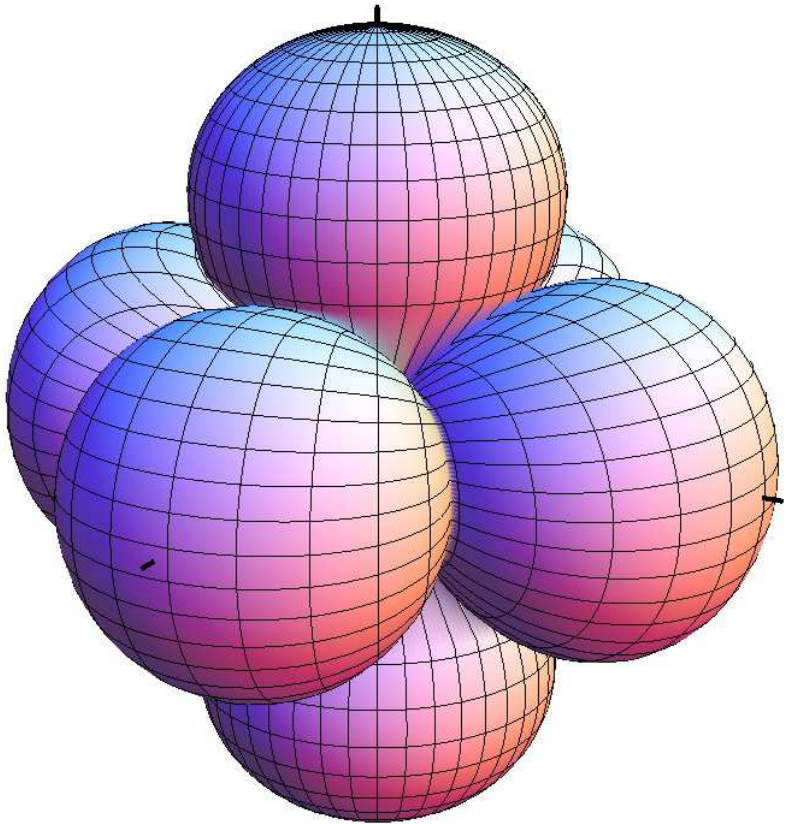,height=12cm} \caption{A plot of the
  flat space tunneling rate (\ref{gamlast}), as a
  function of angular coordinates in three dimensions.  The tunneling
  rate is represented here as being proportional to the distance from the
  origin.}
\label{gamplot}
\end{figure}

The flat space tunneling rate is shown graphically, as a function of angular
coordinates, in figure \ref{gamplot}.
It can be seen that $\kappa_{ij}$ is significantly larger near the axes of
this space, even in these few dimensions.  In more dimensions, the
contrast between maximum and minimum is increased.  This
expression for $\kappa_{ij}$ will be used below to show that even a single
transition (the last one) can produce considerable clustering of probability around
the axes of the flux space.  As the radius of the sphere being
considered (and hence the energy level) is increased, gravitational
corrections must increasingly be taken into consideration.  The effect
of gravity is to reduce the contrast between maxima and minima,
smoothing the distribution of $\kappa_{ij}$ and producing a plot that looks
more like a sphere.

As well as nucleating branes of single charge, one may also wish to
consider the possibility of nucleating stacks of branes
simultaneously.  A study of such events has been carried out by
Garriga and Megevand \cite{Garriga}.  In this work it is concluded that the
instantons for nucleating stacks of branes may well not exist, but
if they do then the tunneling rates for such events will be
approximately given by $B^{(k)}_{ij} \simeq k B_{ij}$, where $k$ is
the number of branes in the stack and $B^{(k)}$ is
the exponent of the tunneling rate for nucleating the stack.  This
formula was found in the limit that the bubble wall is
much less than the Hubble rate of the parent vacuum, in which case the
thin-wall approximation can be  used.  When a sufficiently large
stack is nucleated (again, if this is possible), then the thin wall
approximation is violated and the bound on the minimum possible
tunneling rate \cite{kklt} is saturated.  In this limit, the
exponent of the tunneling rate was found to be $B^{(k)}_{ij} \simeq
\frac{8 \pi^2}{\Lambda_i}$.  This rate is
independent of the everything except the energy of the parent
vacuum, and so once it is met all unlikely tunneling events
proceed with this minimum probability, if instantons for these
transitions exist at all.

The possibility of nucleating stacks of branes significantly alters
the dynamics of inflation, if it is allowed to occur. We consider it
beyond the scope of the present study to perform a detailed analysis
of this phenomenon here.  Instead, we will first proceed using the
standard lore of allowing only transitions between nearest neighbors
in the BP landscape.  After analyzing this situation we will then
reflect on how nucleating multiple branes simultaneously will change
this picture.  We leave it to future studies to determine how likely
these events really are.

\subsection{Anthropic Considerations}

In volume weighted measures of inflation the majority of the 3-volume
of the universe, and the majority of new bubbles formed, are at very
high energies.  However, no observers can exist in these vacua.  Firstly, the
famous anthropic bound on the cosmological constant tells us that
anthropically acceptable vacua must
lie within the range $0 \lesssim \Lambda \lesssim 10^{-120}$ in order
for structure to be able to form \cite{ccbound} (we neglect the possibility of small
negative cosmological constant \cite{ads}, for simplicity).  Secondly, these
high energy vacua decay far too quickly for structure to form.  If we are
interested in the likely locations in the landscape for observers to
find themselves we should therefore not ask simply what are the most
likely vacua, but instead what are the most likely vacua that (a) have
a suitably low cosmological constant, and (b) are suitably
long-lived\footnote{There are a number of other anthropic bounds that
  should also be applied from, for example, the fine structure
  constant and Newton's constant.  For simplicity, and due to the
  limited nature of our model, we will not consider all of these
  bounds here.}.

\subsubsection{Cosmological constant bound}

As we mentioned in the preceding subsection, the condition (a) is
tantamount to considering only vacua within a thin shell of width
$10^{-120}$, and radius $\sqrt{2 \Lambda_0}$.  Any points further from
the origin than this have $\Lambda > 10^{-120}$, and so do not allow
for the formation of structure.  Any points closer to the origin are
collapsing AdS vacua.  The volume, $V$, of this shell is then given by
\be
\nonumber
V= \frac{2 \pi^{J/2}}{(\frac{J}{2}-1)! J} \left[ (\sqrt{2}
  \sqrt{\Lambda_0+10^{-120}})^J-(\sqrt{2} \sqrt{\Lambda_0})^J\right]
\simeq \frac{10^{-120} (2\pi)^{J/2} \Lambda_0^{J/2}}{(\frac{J}{2})!}.
\ee
The density of points in the flux space is $(\prod_i^J{q_i})^{-1}$,
so that the condition for more than one point in this anthropic shell is
\be
\label{(a)}
\frac{10^{-120} (2\pi)^{J/2} \Lambda_0^{J/2}}{(\frac{J}{2})! \prod_i^J
  q_i} >1.
\ee

\subsubsection{Life-time bound}

The condition (b) requires that the time scale $\tau_{ij} =
\kappa_{ij}^{-1}$ is less than the time required for structure to
form, and life to evolve. This is clearly a very difficult time to
specify precisely.  For definiteness we will require the bound that the
vacuum should have a lifetime $\; \gtrsim 10 Gyr$, corresponding to
\begin{equation}
\nonumber \kappa_{ij} \lesssim e^{-138}.
\end{equation}
The location of this boundary could be shifted, depending on the time
at which observers are deemed to appear in any particular theory. 
Transitions out of any point that satisfies condition (a) will be
adequately described by the flat space rate, (\ref{gamlast}).  This
rate is at a minimum along the `diagonal', where $J n_D^2 q_i^2 =
2\Lambda_0$. The condition for suitable stability of these points is
given by \be \label{(b1)} q_i J^{3/2} \gtrsim 13 \Lambda_0^{3/2}.
\ee This condition on the slowest transitions rates is the
requirement that there exist \textit{any} points that live longer
than $10Gyr$.  The condition that \textit{all} points in the
anthropic shell are this stable requires a condition on the fastest
transition rate, at the poles. Here $n_P^2 q_i^2 = 2\Lambda_0$, and
the tunneling rate (\ref{gamlast}) gives \be \label{(b2)} q_i
\gtrsim 13 \Lambda_0^{3/2}. \ee

The anthropic conditions (a) and (b) therefore give us acceptable
ranges of the parameters $q_i$ and $\Lambda_0$, in terms of the
number of dimensions $J$.  Strictly, these relations should not be
considered independent of each other as we require not only a
minimum of one point in the anthropic shell and one region in this
shell where points are stable, but one stable point in the shell.
The combination of conditions (\ref{(a)}) and (\ref{(b1)}) can then
be considered a necessary condition, and the combination of
conditions (\ref{(a)}) and (\ref{(b2)}) a sufficient one. These conditions are summarized in the table below,
where it is assumed that all charges are approximately equal (so
that all $q_i \simeq q$).
\begin{table}[ht]
\begin{center}
\begin{tabular}{llll}
\hline
 & some points \hspace{20pt} &  all points \hspace{20pt} & enough points \\
 & stable &  stable & in shell \\
\hline
J \hspace{30pt} & $q \Lambda_0^{-3/2} \gtrsim$ &  $q \Lambda_0^{-3/2} \gtrsim$ &
$q^{-1} \Lambda_0^{1/2} \gtrsim$ \\
\hline
100 & \qquad 0.013 & \qquad 13 & \qquad 28 \\
250 & \qquad 0.0033 & \qquad 13 & \qquad 8.3 \\
500 & \qquad 0.0012 & \qquad 13 & \qquad 6.7 \\
1000 & \qquad 0.00041 & \qquad 13 & \qquad 7.2 \\
2000 & \qquad 0.00015 & \qquad 13 & \qquad 8.8 \\
\hline
\end{tabular}
\end{center}
\end{table}

It can be immediately seen, for reasonable parameters, that the
condition for all points in the anthropic shell to be stable is often
not satisfied.  This means that the regions near the poles of this
shell are not long-lived enough for observers to evolve.  It can
also be seen that for fixed $\lambda_0$ (\ref{(a)}) provides an
upper bound on $q$, and (\ref{(b1)}) provides a lower bound.
Similarly, for fixed $q$ (\ref{(a)}) provides a lower bound on
$\Lambda_0$, whilst (\ref{(b1)}) provides an upper bound.

We expect the bare cosmological constant to be of order unity, and the
charges $q$ to be a couple of orders of magnitude below
this scale.  Reasonable values for the dimensionality of the phase
space are $100$ to $1000$.  For definiteness in what follows, we will
often consider a model in which $q=10^{-2}$, $\Lambda_0=1/2$ and
$J=500$.  This satisfies the conditions (\ref{(a)}) and (\ref{(b1)}),
but not (\ref{(b2)}).  The condition (\ref{(b2)}) could be satisfied
by considering a smaller $\Lambda_0$. 

\section{Populating at High Energies}

One of the benefits of volume weighted measures of inflation is that
there exists a stable late-time distribution, which can be reached from
a broad range of initial conditions:  Once this `stationary' regime is achieved,
the initial conditions of the system are quickly forgotten.  All that
is required for the universe to achieve stationarity is that at least
one part of it should reach the highest energy states.  Although this
may require unlikely upwards tunneling events, the
rapid expansion of the new higher-energy vacua produced can often be
more than enough to offset the improbability of such events occurring.

Of course there exists an upper boundary beyond which the
semi-classical methods we employ cannot be usefully applied.  This
is the Planck boundary, beyond which a full quantum theory of
gravity is required for an adequate description of the space-time
(if this concept survives).  We will not consider such regimes here,
but will instead restrict ourselves to considering only
sub-Planckian regions. We will now argue that on high energy shells,
of constant $\Lambda$, probability will be distributed approximately
isotropically.

In a landscape with many high energy vacua the evolution of the
universe, according to volume weighted measures, will not be
dominated by a single high-energy vacuum, as may initially be suspected, but rather by the combined
expansion of the many highest energy vacua\footnote{We thank Andrei
  Linde for making this point clear to us.}.  The effect of
the expansion of these vacua will give an exponentially growing
probability
\begin{equation}
\nonumber P_h \sim e^{3 H_h t},
\end{equation}
which will be shared by the lower energy vacua in the manner
outlined in section 2.  Here $H_h$ is the effective Hubble rate of
the highest energy vacua, which we expect to be near Planckian.  We
may now ask ourselves what the condition is for a particular vacuum
to find itself in some `smeared out' region at high energy, where
all $P \simeq P_h$.  To answer this consider the evolution equation
\begin{align*}
\dot{P}_i &= \sum_j \kappa_{ji} P_j - \sum_k \kappa_{ik} P_i + 3 H_i
P_i\\
\simeq 3 H_h P_i &\simeq \sum_j \kappa_{ji} P_h - \sum_k \kappa_{ik}
P_i +3H_i P_i.
\end{align*}
In going from the first to second line we have assumed that all
neighboring vacua to $P_i$ have approximately $P_j \simeq P_h$,
i.e. that our chosen vacuum is in the smeared out region.  This
assumption will be consistent only if $P_i \simeq P_h$, in which case
the vacuum can consistently be described as being in a `smeared out'
region at high-energy.  To see if this is true we can rearrange the
equation above to obtain
\be
\nonumber
\frac{P_h}{P_i} \simeq \frac{3 \Delta H+ \sum \kappa_{i
    \downarrow}}{\sum \kappa_{\downarrow i}},
\ee
where $\Delta H \equiv H_h-H_i$.  Now, if $3 \Delta H \lesssim \sum
\kappa_{i \downarrow}$, then
\be
\nonumber
\frac{P_h}{P_i} \simeq \frac{\sum \kappa_{i \downarrow}}{\sum
  \kappa_{\downarrow i}} \sim O(1).
\ee
In this case we have $P_i\sim P_h$, and the vacuum under
consideration is in the smeared out region.  (The order of magnitude
here is due to the small difference in energy levels between
neighboring high-energy vacua in the BP landscape).  Alternatively, when
$3\Delta H \gg \sum \kappa_{i \downarrow}$, we have
\be
\nonumber
\frac{P_h}{P_i} \simeq \frac{3 \Delta H}{\sum
  \kappa_{\downarrow i}} \gg 1,
\ee
and $P_i \ll P_h$.  In this case the vacuum we are considering cannot
consistently be assumed to be in the smeared out region.

Furthermore, transitions between the highest energy vacua in the BP
landscape will be only very weakly suppressed, with $\kappa_{ij}
\sim O(1)$.  As such, the value of $\kappa_{ij}/3H_h$ will also be $O(1)$
and the ratio of probabilities given by (\ref{gentun}) will be
\be
\nonumber
\frac{P_j}{P_h} \sim O(1).
\ee
The products of any such decay will then be (approximately) as equally abundant
as the vacua from which they decayed.  If the highest energy vacua
decay at these rates then the whole of the
high-energy part of the landscape will very quickly become populated
with an approximately flat distribution of probability.

We have identified here reason for thinking that the highest
energy levels of the BP landscape should have a probability
distribution that is `smeared out' (i.e. isotropically
distributed).  This smearing process erases any anisotropy, or favoring
of any particular point or energy level in the initial conditions of
the universe -- as long as the highest energy vacua can be reached, we
are left with a landscape that is populated at high energies by an
approximately flat distribution of probability.  Clearly this
is a very different regime to the low energy one, in which tunneling
events are suppressed by large factors of $3 H_h/\kappa_{ij} \gg 1$:
Hence, there exists a boundary above which we can expect tunneling rapid
enough to produce a flat probability distribution, and below which
tunneling is heavily suppressed.

\section{Tunneling from High to Low Energies}\label{tunnelingdown}

We will now consider how tunneling from high energies down to lower
energies will proceed. First it is necessary to determine which
paths will contribute the highest probability flux onto the
anthropic shell.  The shortest path from any point on the anthropic
shell to the `smeared out' surface, described above, will be a
straight line extending in one dimension of the flux space.  Any
deviations from this shortest path will add extra factors of
$\kappa_{ij}/3H_h$, which will suppress their relative contribution
to the overall flux arriving at our point.  It is now necessary to
determine whether this suppression will be sufficient to beat the
combined contribution of the potentially large number of paths which
are slightly longer, and less favorable.

The degree to which a longer path can be considered less favorable
depends upon where it deviates from the straight line, shortest
path, and whereabouts in the flux space the shortest path is
located.  For example, a straight-line shortest path which is
parallel, and close to an axis of the flux space will have very
little contribution from any other (longer) paths.  This is due to
the vacua along such a trajectory having one large flux, and the
rest very small. Transitions in directions with small fluxes
associated with them will then only progress at very low rates,
compared to those in the direction of the single large flux.  The
shortest path, which is oriented in the direction of the large flux,
will then consist of a series of rapid transitions.  Any longer path
will necessarily include additional transitions in a direction
associated with a smaller flux, and so will be heavily suppressed.
Away from the axes extra transitions at or near the surface beyond
which rates are no longer suppressed will not incur a large penalty.
We will treat this high energy surface as an initial boundary from
which tunneling down to the anthropic shell begins. Any unsuppressed
transitions before this surface can then be considered as
contributing to the boundary, but not favoring any particular path
thereafter.  We will show that our results are not heavily dependant
on the exact location of such a boundary.  Extra deviations away
from the shortest path at low energies will incur a large penalty,
which will be enough to make any such contribution
to the total probability negligible.  We therefore expect the
shortest path from the high-energy boundary to be a good
approximation to the probability of arriving at an acceptable point
in the anthropic shell, both near and far from the axes of the phase
space.

We now need the appropriate transition rates along these shortest
paths. Although the approximation to the gravitational correction
(\ref{r1}) is the appropriate one for most of the high-energy
transitions (in a high dimensional landscape), it is not the one in
which we are interested here.  A straight-line extending on one
dimension of the flux space, from the anthropic shell to high
energies, will always be contained within a cylinder about one of
the axes, of cross-sectional area less than that of the anthropic
shell itself.  The appropriate gravitational correction is therefore
given by (\ref{r2}) along such a path, until the last transition
where it is given by (\ref{gamlast}).  In this section we will
consider the regime where (\ref{r2}) is valid.  The effect of the
last transition will be considered separately in the next section.

\begin{figure}[t]
\center \epsfig{file=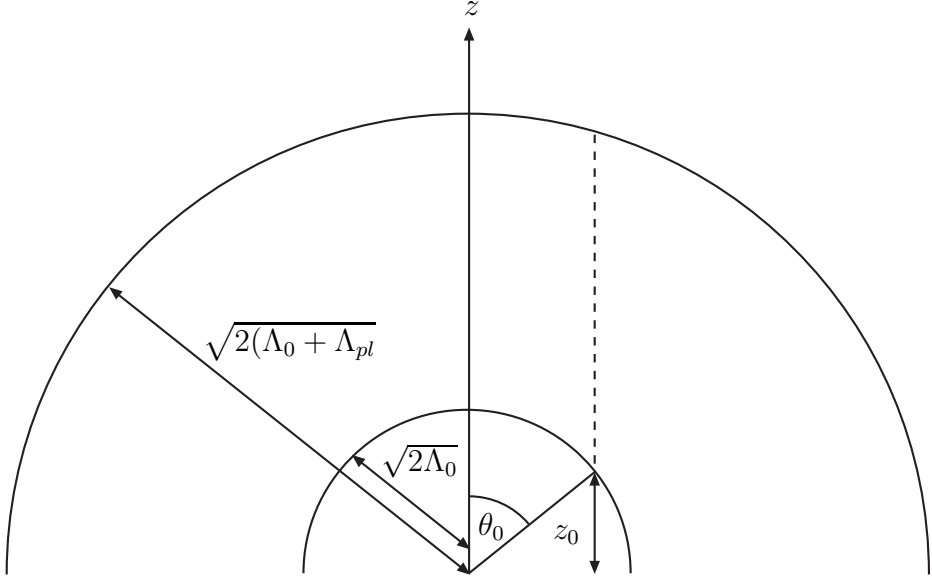,height=8cm} \caption{A schematic of the
shortest path from high energies to the penultimate sphere, parallel to the
$z$ axis.}
\label{short}
\end{figure}

The tunneling rate prescribed by (\ref{r2}) is
\be
\nonumber
\kappa_{ij} \simeq \textrm{Exp}\left[ -\frac{27 \pi^2}{8 n_i^3 q_i^2} \left\{ \frac{4
    n_i^2 q_i^2-3 \Lambda_i}{4 n_i^2 q_i^2+3 \Lambda_i} \right\}
   \right]
\ee
and the total probability of tunneling down from high energies to the
penultimate sphere, before the anthropic one, is then given by the
product of these tunneling rates along the shortest path, shown
schematically in figure \ref{short}.

\psfrag{2L0PL}{$\sqrt{2(\Lambda_0+\Lambda_{pl}}$}
\psfrag{2L0}{$\sqrt{2\Lambda_0}$} \psfrag{th}{$\theta_0$}
\psfrag{z}{$z$} \psfrag{z0}{$z_0$}

The value of $\Lambda_i$ along this path is
\be
\label{L}
\Lambda_i = \frac{1}{2} n_i^2 q_i^2-\frac{1}{2} z_0^2,
\ee
where $z_0^2 \equiv 2 \Lambda_0 \cos^2 \theta_0$ and $\theta_0$ is
shown in figure \ref{short}.  The value of $n_i q_i$, for any $i$, is
\be
\label{nq}
n_i q_i = q_i (n_{0i}+j) = z_0 +j q_i,
\ee
where $n_{0i}$ is the value of $n_i$ in the anthropic shell, and $j$ is
the number of flux quanta in $z$ away from that point.  Now the total
supression along this path is given by
\be
\nonumber
\kappa_T \equiv \prod_{i}^J \kappa_{i \downarrow} = \textrm{Exp} \left[- \frac{27 \pi^2
    q_i}{8} \sum_{j=2}^N \left( \frac{4 n_i^2 q_i^2-3 \Lambda_i}{4 n_i^2
    q_i^2+3 \Lambda_i} \right) \frac{1}{n_i^3 q_i^3} \right]
\ee
where $\Lambda_i$ and $n_i q_i$ are given by (\ref{L}) and (\ref{nq}),
and $N$ is the number of transitions involved in tunneling from high
energy to low energy.  In a continuum approximation this expression
can be approximated by
\begin{align}
\nonumber
\kappa_T &\simeq \textrm{Exp} \left[- \frac{27 \pi^2
    q_i}{8} \int^{N+1/2}_{3/2}\left( \frac{4 n_i^2 q_i^2-3 \Lambda_i}{4
    n_i^2 q_i^2+3 \Lambda_i} \right) \frac{dj}{n_i^3 q_i^3} \right]\\
&\simeq \textrm{Exp} \left[ -\frac{27 \pi^2}{8} \left[ \frac{8\ln \left\{
    \frac{6\Lambda_i}{n_i^2 q_i^2}+8\right\}}{3 (n_i^2 q_i^2-2 \Lambda_i)}
    +\frac{1}{2 n_i^2 q_i^2}\right]_{3/2}^{N+1/2} \right].
\label{oops}
\end{align}
As we have already discussed, the leading contribution to this
quantity will be the low energy limit, where extra transitions are
heavily suppressed.  The contribution from extra transitions at high
energy will be negligible, and correspondingly we take $N \rightarrow \infty$.
In this limit we then have
\be
\nonumber
\kappa_T \simeq \textrm{Exp} \left[\frac{27
    \pi^2}{8} \left( \frac{8}{3z_0^2} \ln \left(1-\frac{12
    z_0^2}{11 (2z_0+3q_i)^2}\right)+\frac{2}{(2z_0+3 q_i)^2}\right) \right].
\ee
The limit $N \rightarrow \infty$ is further justified in appendix \ref{A1},
where integrations are performed to finite values of $N$.  It is found
that the quantities we are interested in are largely insensitive to the
exact location of the high-energy boundary, and so we are justified in
approximating it to be at infinity.  For $z_0 \gg q_i$ the expression above becomes
\be
\kappa_T \simeq \textrm{Exp} \left[- \frac{27
    \pi^2}{8 z_0^2} \left( \frac{8}{3} \ln \left( \frac{11}{8}\right)
  -\frac{1}{2}\right)\right] \simeq \textrm{Exp} \left[-\frac{12}{z_0^2} \right].
\label{highlow} \ee 
It can be clearly seen that the tunneling rate
is significantly faster to points in the anthropic shell with large
$z$ (i.e. close to the axes, away from the diagonal).  This is due to
the supression associated with the extra transitions required to get to
the points further from the axes.  We will now
use this expression for the tunneling rates to determine where on
the sphere a `typical' vacuum is located.  In order to do this we
will need to balance the faster transition rates at the poles of the
sphere, with the increased volume along the diagonals.

\subsection{Expectation Values}

In order to consider the distribution of probability on a sphere it
is useful to determine the moments
\begin{equation}
\langle f^2\rangle \equiv\int f^2\kappa_T d \Omega
\end{equation}
where $f=f(x_i,z)$ is any function of $x_i$ and $z$ on the sphere, and
$d \Omega$ is the surface element of a hypersphere.  This
expression takes into account not only the effect of the transition rate
$\kappa_T$, but also the effect of increasing surface area away from
the axes.

We can consider the penultimate sphere as being divided into $2 J$
symmetric sectors, each centered around one of the $2J$ axes of the
space.  Each of these sectors corresponds to the region $\sigma$ on
the surface of the penultimate sphere which is closest to the axis
about which it is centered.  $z$ is then given, in each respective
region, as the corresponding value of the Cartesian coordinate along
its central axis, and the whole sphere is built up by evaluating
$\langle n^2 \rangle $ in any one of the regions and multiplying it
by $2 J$. The integral that we are now required to evaluate is
\begin{equation} \label{Aq}
\langle f^2\rangle = \int_0^1 dz \prod_i^{J-1} \int_{-z}^z dx_i f^2
e^{-12/z^2} \delta \left( \sum_i^{J-1} x_i^2 + z^2 -1
\right),
\end{equation}
where we have taken $\Lambda_0=1/2$.  The Cartesian coordinates $x_i$
correspond to all of those directions which are
orthonormal to the central axis $z$.  The limits of integration,
$-z$ and $z$, then pick out the sector discussed above. Included in
the integrand is a delta function intended to select the surface at
$\sim \sqrt{2 \Lambda_0} = 1$, which corresponds to the anthropic shell.
Here $\langle f^2 \rangle $ is being evaluated in any one of the $2 J$
symmetric regions.  We will be interested in the choices $f=z$ and $f=x$,
where $x$ is any Cartesian coordinate orthonormal to $z$. The
moments $\langle z^2 \rangle $ and $\langle x^2 \rangle $ then give us the expectation values of
these quantities, and their ratio gives us a measure of the peaking
of probability on the anthropic sphere due to tunneling down from high
energy to the penultimate sphere before the anthropic one.

In appendix \ref{appendixA} we find that the expectation values for
$\langle z^2 \rangle$ and $\langle x^2 \rangle$ can be given in terms of
one-dimensional integrals as
\begin{equation}
\langle z^2\rangle=\frac{2}{\Gamma(\frac{J+2}{2})} \int_0^\infty
dy y^2 \left[\frac{\sqrt{\pi}
  y}{\sqrt{12+y^2}} E(\sqrt{12+y^2}) \right]^{J-1} e^{-(12+y^2)}
\label{intq}
\end{equation}
and
\begin{align}\nonumber
\langle x^2\rangle &=\frac{2}{\Gamma(\frac{J+2}{2})} \int_0^\infty dy
 y^2 \left( \frac{1}{2 (12+y^2)}- \frac{e^{-(12+
     y^{2})}}{\sqrt{\pi} \sqrt{12+y^2} E(\sqrt{12+y^2})  } \right)
 \\ &\hspace{200pt} \times \left[\frac{\sqrt{\pi} y}{\sqrt{12+y^2}} E(\sqrt{12+y^2}) \right]^{J-1}
e^{-(12+ y^{2})}. \label{int2q}
\end{align}
Here the function $E(p)$ is the integral of a Gaussian distribution,
or error function, given by $2\pi^{-1/2} \int_0^p e^{-q^2}dq$.

\begin{figure}[t]
\center \epsfig{file=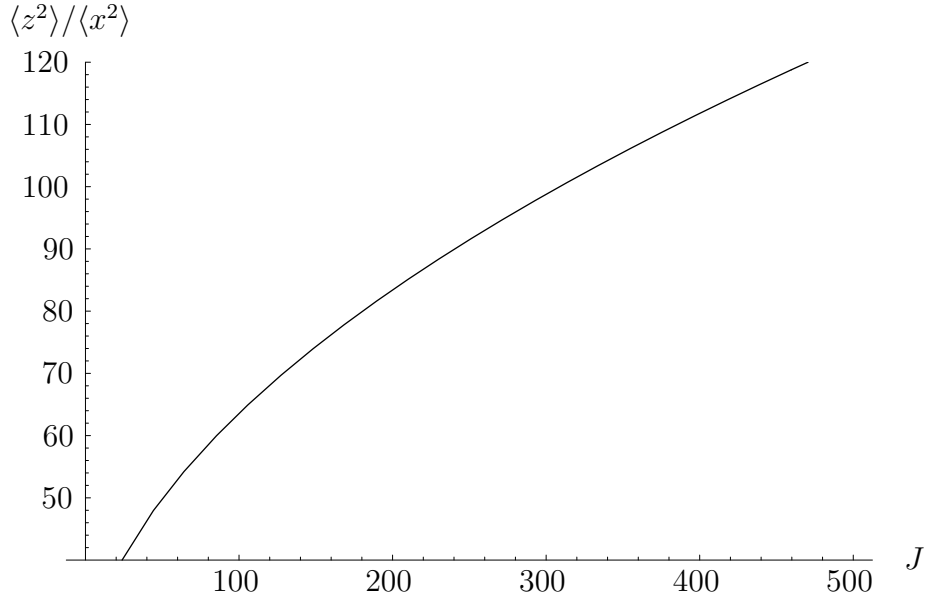,height=8cm} \caption{A plot of
$\langle z^2\rangle / \langle x^2\rangle$ as a function of the dimensionality of the
  flux space, $J$.}
\label{hlplot}
\end{figure}

The measure of peaking that we are interested in is given by the
ratio of these two expressions, $\langle z^2\rangle / \langle
x^2\rangle$. This is shown graphically in figure \ref{hlplot}.  It
can be seen from this plot that the peaking produced by tunneling
down to the penultimate sphere is extremely sharp, and increases
with the dimensionality of the flux space.
\psfrag{jj}{$J$} \psfrag{z2overx2}{$\langle z^2 \rangle/\langle x^2
\rangle$} \psfrag{50}{50} \psfrag{60}{60} \psfrag{70}{70}
\psfrag{80}{80} \psfrag{90}{90} \psfrag{100}{100} \psfrag{110}{110}
\psfrag{120}{120} \psfrag{200}{200} \psfrag{300}{300}
\psfrag{400}{400} \psfrag{500}{500}
A flat distribution of probability would produce the result $\langle z^2 \rangle/\langle
x^2 \rangle = 1$.  The large values of this quantity here show that
there exists a strong peak in probability on the anthropic shell.

\subsection{The Single Peak Approximation}

We are now in a good position to investigate approximations that
will make these quantities more transparent.  A particularly useful
simplification is to extend the domain of integration of equation
(\ref{Aq}), to make it independent of $z$.  If we were to allow the
limits of integration for the coordinates $x_i$ to run from $-1$ to
$1$, instead of $-z$ to $z$, then this would allow us to transform
to a hyper-cylindrical coordinate system in which $J-1$ of the
dimensions could be integrated over in a straightforward fashion.
This would leave a simple one-dimensional integral that can be
solved by saddle point approximation.  We will call this the
single peak approximation, as we will be integrating over a larger
region that should contain several peaks of probability; we will,
however, ignore all but the single central peak of the initial
region and show that this is a good approximation to integrating
over the correct domain, as in (\ref{Aq}).  The reason why this is a
plausible approximation to consider is that the rate of transitions
(\ref{highlow}) can be seen to become very small at low $z$.  The
extra region being integrated over in the single peak approximation is
all at low $z$, and so it is a reasonable proposition to ask if
integrating over such an extended domain yields similar results to
integrating over the actual domain of interest.  We will argue that
it does.

The single peak approximation yields the results
\begin{align}
\nonumber \langle\bar{z}^2\rangle &=
\frac{2}{\Gamma(\frac{J+2}{2})} \int_0^\infty dz \prod_i^{J-1}
\int_{-\infty}^{\infty} dx_i z^2 e^{-(\sum_i x_i^2 +z^2)(1+12 z^{-2})}\\
&= \frac{2}{\Gamma(\frac{J+2}{2})} \int_0^\infty dz z^2
\left[\frac{\sqrt{\pi}
  z}{\sqrt{12+z^2}} \right]^{J-1} e^{-(12+z^2)},
\end{align}
which can be compared to equation (\ref{intq}), and
\begin{align}
\nonumber \langle\bar{x}^2\rangle &=
\frac{2}{\Gamma(\frac{J+2}{2})} \int_0^\infty dz \prod_i^{J-1}
\int_{-\infty}^{\infty} dx_i x^2 e^{-(\sum_i x_i^2 +z^2)(1+12 z^{-2})}\\
&=  \frac{2}{\Gamma(\frac{J+2}{2})} \int_0^\infty dz
 \frac{z^2}{2 (12+z^2)} \left[\frac{\sqrt{\pi}
     z}{\sqrt{12+z^2}} E(\sqrt{12+z^2}) \right]^{J-1}
 e^{-(12+ z^{2})},
\end{align}
which can be compared to equation (\ref{int2q}).  We denote
quantities evaluated in the single peak approximation by an over-bar.
The fractional error of considering this approximation can
then be written as
\begin{equation} \delta \equiv
\frac{\langle z^2\rangle / \langle
x^2\rangle-\langle\bar{z}^2\rangle / \langle\bar{x}^2\rangle}{\langle
z^2\rangle / \langle x^2\rangle}.
\end{equation}
This quantity is vanishingly small for all $J$, showing that the single
peak approximation is very good.

\subsection{Analytic Approximations}

Using the single peak approximation we will now find analytic
expressions for these moments. Recall the
tunneling rate is given by
\begin{equation}
\kappa_T\sim \textrm{Exp}\left[-\frac{12}{z^2}\right] =
\textrm{Exp}\left[-\frac{K}{z^2}\right],
\end{equation}
where $K=12$, and to be concise we have replaced $z_0$ with $z$.  The
normalization of the probability distribution is then
\begin{equation}
N=\int\kappa d\Omega = 2J\int_0^1dz\prod_i^{J-1}\int_{-1}^1 dx_i
e^{-K/z^2}\delta\left(\sum_j^{J-1}x_i^2+z^2-1\right)\ .
\end{equation}
In the single peak approximation, the limits on the $x_i$ integrals
are replaced with -1 and 1.  Changing coordinates from Cartesian to
hyper-cylindrical, and defining $r^2=\sum_i^{J-1}x_i^2$, we can rewrite $N$ as
\begin{equation}
N=2JV_{J-2}\int_0^{1}dz\int_0^{1}dr
r^{J-2}e^{-K/z^2}\delta(r^2+z^2-1)\ ,
\end{equation}
where $V_{J-2}$ is the volume of a $(J-2)$-sphere. Eliminating
the delta function with the $r$-integral leaves us with an integral over $z$:
\begin{equation}
\label{N2} N=2JV_{J-2}\int_0^{1}dz
(1-z^2)^{\frac{J-3}{2}}e^{-K/z^2}\sim2JV_{J-2}\int_0^{1}dz
e^{-\left(K/z^2-\frac{J}{2}\textrm{ln}(1-z^2)\right)}\ .
\end{equation}
This integral can be approximated by a saddle point, which is at
\begin{equation}
z^2=\frac{K}{J}\left(-1+\sqrt{1+\frac{2J}{K}}\right)\sim\sqrt{\frac{2K}{J}}
=\left(\frac{24}{J}\right)^{1/4}.
\end{equation}
At first glance it may seem that this result is in contradiction
with the validity of the single peak approximation. After all, $z$
is small for any reasonable $J$ ($\tilde{C}\lesssim 1$) and
therefore $r$ is close to 1. However, although $z$ is small, a
typical $x_i$ ($\sim 1/\sqrt{J}$) is much smaller and so the peak in
the distribution on the sphere is, in general, closer to the
$z$-axis than to any of the other $x_i$ axes. This issue is
discussed in more detail in section \ref{probable}.

We can similarly calculate the
expectation values of powers of the fluxes:
\begin{equation}
\label{expected1} \langle z^k\rangle=\frac{1}{N}V_{J-2}\int_0^{1}dz
z^k(1-z^2)^{\frac{J-3}{2}}e^{-K/z^2}\sim
\frac{1}{N}V_{J-2}\int_0^{1}dz
e^{-\left(K/z^2-\frac{J}{2}\textrm{ln}(1-z^2)-k\,\textrm{ln}(z)\right)}\
.
\end{equation}
This should be normalized using the $N$ calculated
above. The integrand is again extremized at $z\sim
\left(\frac{2K}{J}\right)^{1/4}$ (with an $O((2K/J)^{1/4}(k/J))$
correction). Substituting this back into the integral above,
and evaluating the saddle point, gives us
\begin{equation}
\langle z^k\rangle\sim \left(\frac{2K}{J}\right)^{k/4}=  \left(\frac{24}{J}\right)^{k/4}\ .
\end{equation}
Here the normalization $N$ cancels the exponential, the remaining
integral and various other numerical factors.  In a flat measure the same expectation value is given by
\begin{equation}
\label{zkk2} \langle
z^k\rangle_{\textrm{flat}}=\frac{V_{J-2}}{N_{\textrm{flat}}}
\int_0^{\infty}dr (1-r^2)^{(k-1)/2}r^{J-2}\ .
\end{equation}
The scaling with $J$ can again be obtained, and is now found to be
\begin{equation}
\langle z^k\rangle_{\textrm{flat}}\sim\left(\frac{1}{J}\right)^{k/2}.
\end{equation}
Symmetry considerations then give the expected values of the remaining
fluxes as
\begin{equation}
\langle x_i^k \rangle\sim \frac{1}{J^{k/2}},
\end{equation}
and a typical flux vector can be estimated as
\begin{equation}
\label{zkk3} q \vec{n}\sim\left(\left(\frac{2K}{J}\right)^{1/4}\ ,\
\frac{1}{J^{1/2}}\ ,\ \ldots\ ,\ \frac{1}{J^{1/2}}\right)\ ,
\end{equation}
which, for $J=500$ and $q=10^{-2}$, is
\begin{equation}
\vec{n}\sim\left(45\ ,\ 4.5\ ,\ \ldots\ ,\ 4.5\right)\ .
\end{equation}
Therefore, the most likely vacua produced by tunneling down from
high energies to the penultimate sphere are those with one flux
approximately ten times as large as the others.  The spread of
probability about this most likely point will be addressed in
section 6.  Here we have assumed that all charges are the same.
Allowing different charges will produce sharper peaking near the axes
associated with smaller $q$.

\section{Transitions into the Anthropic Shell}\label{finalhop}

Having calculated the probability of transitioning down onto the
penultimate sphere, above the anthropic one, we will now evaluate the
effect of the last jump.  The final transition rates into the
anthropic shell are well approximated by the flat space rates
\begin{equation}
\label{tran}
\kappa_F \sim e^{-\tilde{C}/z^3}
\end{equation}
where $z$ denotes the value of the Cartesian coordinate in the
direction of transition, and $\tilde{C}$ is a constant ($\tilde{C}\simeq 0.3$,
when $q_i=10^{-2}$).  Strictly, $z$ should be a discrete number and correspond to the
penultimate shell, one step up from the anthropic one.  In our
continuum approximation, however, $z$ is a continuous variable,
which we can evaluate on the surface of the anthropic shell.

Transitions will occur in all $J$ dimensions of the space. However,
we see that $\kappa_F$ is very strongly dependent on the value of
$z$, allowing us to approximate the probability density on the
anthropic sphere as being primarily due to transitions from the
largest of these $z$.  We again consider the sphere as
being divided into $2 J$ symmetric sectors. The integral that we are
now required to evaluate is
\begin{equation} \label{A}
\langle f^2\rangle = \int_0^1 dz \prod_i^{J-1} \int_{-z}^z dx_i f^2
e^{-\tilde{C}/z^3} \delta \left( \sum_i^{J-1} x_i^2 + z^2 -1
\right).
\end{equation}
The Cartesian coordinates $x_i$ correspond to all of those which are
orthonormal to the central axis $z$.  The limits of integration,
$-z$ and $z$ pick out the sector of interest.

In appendix \ref{appendixB} we show that the single peak
approximation for the final transition is accurate to better than
$0.03 \%$ at $J = 500$, and that this accuracy increases as $J$
is increased. Having justified the validity of this approximation, we
will now use it to obtain analytic expressions for the peaking of
probability produced by the final transition.

\subsection{Peaking From the Last Step}

Working in the single peak approximation we can use saddle points
methods, as before, to estimate the level of peaking from
the final transition. We begin with the integral
\begin{equation}
N=\int\kappa d\Omega = 2J\int_0^1dz\prod_i^{J-1}\int_{-z}^zdx_i
e^{-\tilde{C}/z^3}\delta\left(\sum_j^{J-1}x_i^2+z^2-1\right)\ ,
\end{equation}
which has its integrand extremized at
\begin{equation}
\label{43}
z \sim \left(\frac{3\tilde{C}}{J-2}\right)^{1/5}\ .
\end{equation}
Again, the integral has most of it's support relatively close to the
$z$ axis, consistent with the single peak approximation.

\begin{figure}[t]
\center \epsfig{file=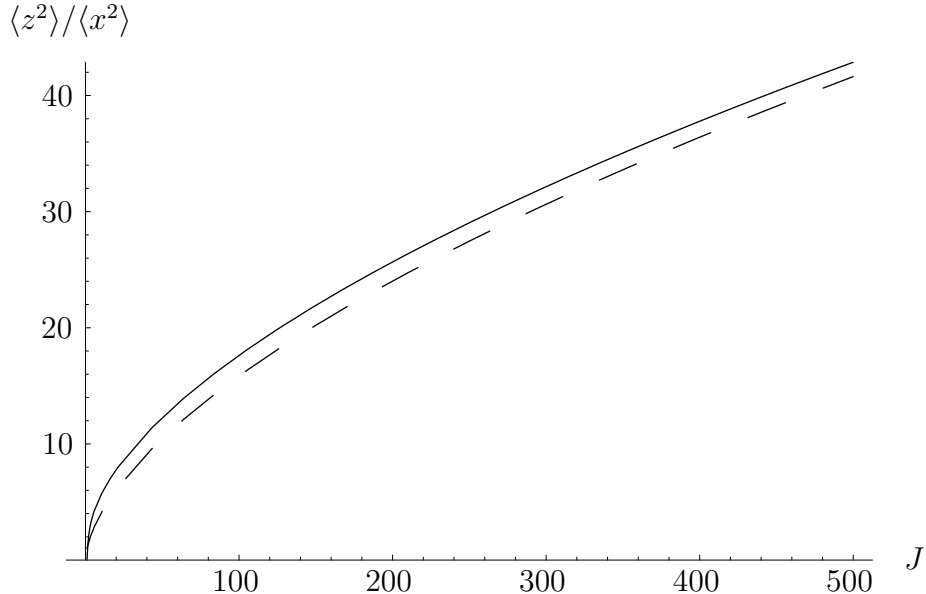,height=8cm}
\caption{A plot of
  $\langle z^2 \rangle /\langle x^2 \rangle$ calculated numerically
  (solid line), and by saddle point
  approximation (dashed line).  The saddle point approximation is
  taken from (\ref{zk4}), and can be seen to be a reasonable
  approximation to the numerics at large $J$.  The accuracy of the
  saddle point approximation increases with $J$.}
\label{z2}
\end{figure}

Once again we can calculate the expected fluxes from the integral
\begin{equation}
\label{zk} \langle z^k\rangle=\frac{V_{J-2}}{N} \int_0^{\infty}dr
(1-r^2)^{(k-1)/2}r^{J-2}e^{-\tilde{C}/(1-r^2)^{3/2}}\ ,
\end{equation}
which has its saddle point given by \ref{43} (with an
$O(\epsilon k^2/J)$ correction), and thus
\begin{equation}
\langle
z^k\rangle\sim\left(\frac{3\tilde{C}}{J-2}\right)^{k/5}\sim\left(\frac{1}{J}\right)^{k/5}.
\end{equation}
This allows us to write a typical vector as
\begin{equation}
\label{zk4}
q \vec{n}\sim\left(\frac{1}{J^{1/5}}\ ,\ \frac{1}{J^{1/2}}\ ,\ \ldots\
,\ \frac{1}{J^{1/2}}\right)\ .
\end{equation}
For $J=500$ and $q = 10^{-2}$ we should then typically expect to find
one set of preferred fluxes to be over 6 times greater than the others:
\begin{equation}
\vec{n}\sim\left(29\ ,\ 4.5\ ,\ \ldots\ ,\ 4.5\right)\ .
\end{equation}
These saddle point approximations are shown graphically in figure
\ref{z2}, where they are compared to direct numerical
integrations.

\psfrag{10}{10} \psfrag{20}{20} \psfrag{30}{30} \psfrag{40}{40}

The peaking found here is clearly less than that found in the previous section,
and we will now show that its effects are negligible in comparison.
The total tunneling rate from combining (\ref{highlow}) and
(\ref{tran}) is
\be
\nonumber
\kappa \simeq \text{Exp} \left[ -\frac{12}{z_0^2}-\frac{0.3}{z_0^3} \right].
\ee
It can be seen that the second term in this expression only becomes
important at about $z_0 \sim 1/40$.  Before this the first term
dominates, and the expression is well approximated by (\ref{highlow}).
This critical value of $z_0$ can be seen to be an order of magnitude
smaller than the location of the saddle points involved in either
this, or the previous, section.  As both of the saddle-points are
located comfortably in the region where (\ref{highlow}) is a good
approximation to the total tunneling rate, we consider the effects of
this last transition to be negligible in comparison to those
considered in the last section.  This is supported by the relatively
mild peaking found above.

\section{The Most Probable Vacua}\label{probable}

Having demonstrated that regions of the anthropic shell are
prefentially selected by
volume-weighted measures, we now turn our attention to how sharp
this peaking is.  We begin by finding
analytic estimations for the width of the probability
distributions, and then use these to obtain characteristic numbers for
the volume of the universe that exists in some small fraction
of the anthropic sphere.  We then consider the issue
of life-times.  By imposing an anthropic limit on the minimum
life-time of acceptable vacua we find that there exists
a pressure favoring shorter life-times.

\subsection{Peaking About the Most Likely Point}

We will now refine our assertion that the
measure on the anthropic sphere is sharply peaked, and give a
demonstration of the level of this peaking. We begin our argument by
considering the tunneling down from high to low energies, as discussed
in section \ref{tunnelingdown}.  We do this by noting that near the peak the
distribution is Gaussian, and then calculating the standard deviation.

\begin{figure}[t]
\center \epsfig{file=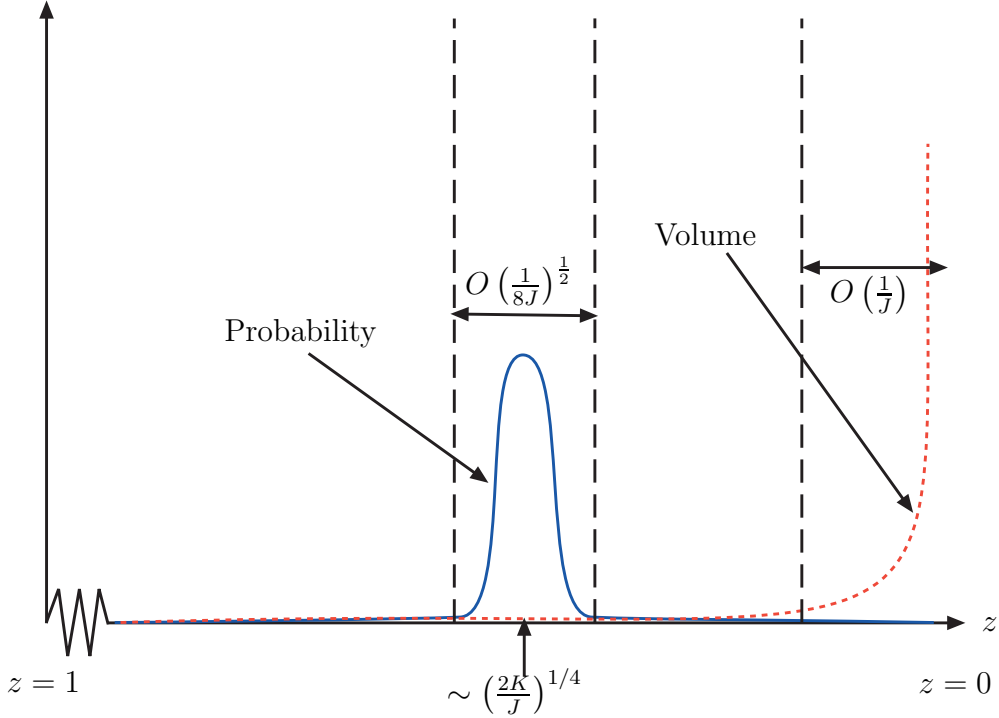,height=10cm} \caption{A sketch
of the peaking of the probability measure as a function of $z$,
along with the volume of the sphere.  The overlap between the
probability peak and the volume peak decreases as $J$ is increased.}
\label{peakingvolume}
\end{figure}

We begin with the integral (\ref{N2}) and note that the saddle point
approximation gives us not only the position of the peak, but also
the width. The leading contribution to this is found to be
\begin{equation}
\sigma^2\sim\frac{1}{8J}.
\end{equation}
Now consider the flat probability distribution on the sphere, where
\begin{equation}\label{flatvolume2}
N_{\textrm{flat}}=V_{J-2}\int dz(1-z^2)^{(J-3)/2},
\end{equation}
at $z\ll 1$ this can be approximated by
\begin{equation}\label{flatvolume3}
N_{\textrm{flat}}\sim V_{J-2}\int dz\textrm{Exp}[-Jz^2/2]\ .
\end{equation}
Evidentally the integrand falls at least exponentially for
$z\gtrsim J^{-1}$, and grows rapidly above $z\lesssim J^{-1}$.
Roughly speaking, this means that whilst the probability
distribution integral has most of its support around
$z^4=24/J$, most of the volume of the sphere
lies further out, around $z=1/J$. This is represented
graphically in figure \ref{peakingvolume}, where it is clear that
the overlap between the probability peak and the volume peak
decreases as $J$ is increased -- i.e. more and more of the volume of
the anthropically acceptable parts of the multiverse come from
smaller and smaller fractions of the surface of the anthropic
sphere.

\psfrag{zx}{$z$} \psfrag{Prob}{Probability} \psfrag{vol}{Volume}
\psfrag{r}{$z=0$} \psfrag{z00}{$z=1$}
\psfrag{o(7/10)}{$O\left(\frac{1}{8J}\right)^{\frac{1}{2}}$}
\psfrag{o(1)}{$O\left(\frac{1}{J}\right)$}
\psfrag{peak}{$\sim\left(\frac{2K}{J}\right)^{1/4}$}

We can then estimate what fraction, $f$, of the sphere is covered by
$m \sigma$ of the probability distribution, by evaluating the volume
integral (\ref{flatvolume3}) between the limits $\langle
z\rangle-m\sigma$ and $\langle z\rangle+m\sigma$:
\begin{equation}
f=2J\frac{V_{J-2}}{V_{J-1}}\int_{\langle
z\rangle-m(8J)^{-1/2}}^{\langle z\rangle+m(8J)^{-1/2}}dz
\textrm{Exp}[-Jz^2/2]\ .
\end{equation}
We can bound this integral by multiplying the maximum of volume
integrand (which is at $z=(2K)^{1/4}/J^{1/4}+m(8J)^{-1/2}$) by the
total range of the integral ($2m\sigma$). The leading contribution
is given by
\begin{equation}
f\sim 2J\frac{V_{J-2}}{V_{J-1}}\frac{2m}{J}e^{-Jz^2/2}\sim
2J\frac{V_{J-2}}{V_{J-1}}\frac{2m}{J} e^{-\frac{(24J)^{1/2}}{2}}\sim
4m\sqrt{J}e^{-\frac{(24J)^{1/2}}{2}}.
\end{equation}
For $J=500$ we find that 99.9999\% of the volume of the universe (5
standard deviations) covers $\sim10^{-23}$ of the anthropic region.

This argument can be repeated for the final transition down onto the
anthropic sphere, discussed in section \ref{finalhop}. Here the
fraction of vacua on the anthropic sphere within $m\sigma$ of the
peak is
\begin{equation}
f\sim 2J\frac{V_{J-2}}{V_{J-1}}\frac{2m}{J^{7/10}}
e^{-\frac{J^{3/5}}{2}}\sim 4mJ^{4/5}e^{-\frac{J^{3/5}}{2}},
\end{equation}
where $\sigma\sim J^{-7/10}$.  For $J=500$ we then find that 99.9999\%
of the volume covers only $\sim10^{-6}$ of the anthropic
sphere. Evidentally, this is considerably less than that due to tunneling down from higher
energies.

\subsection{Pressure Against the Life-Time Boundary}

In the preceding sections we have found the most probable vacua that
have a cosmological constant in the anthropically acceptable range
$0 \lesssim \Lambda \lesssim 10^{-120}$.  However, for observers to
appear we require not only an anthropically acceptable value of
$\Lambda$, but also a vacuum that is long enough lived for life to
evolve.  In section 2.3 this was taken to be $\sim 10$ Gyr.  The
vacua located at the peaks of probability (found in the previous
sections) do not satisfy the bound $\tau > 10$ Gyr, and so are
unlikely to be stable enough for observers of our type to occur.  If
we wish to determine the probability of vacua that have both an
acceptably small $\Lambda$, and an acceptably large life-time, then
we must calculate probabilities only up to the boundary at which
$\tau \simeq 10Gyr$. Using our knowledge that there exists a sharp
peak of probability outside of this region, we should expect to find
a selection pressure pushing against the anthropic life-time barrier
(so that vacua closer to the peak are favored).  Here we will
investigate how strong this selection pressure is, and use it to
make a prediction for the half-life of acceptable vacua.

\begin{figure}[t]
\center \epsfig{file=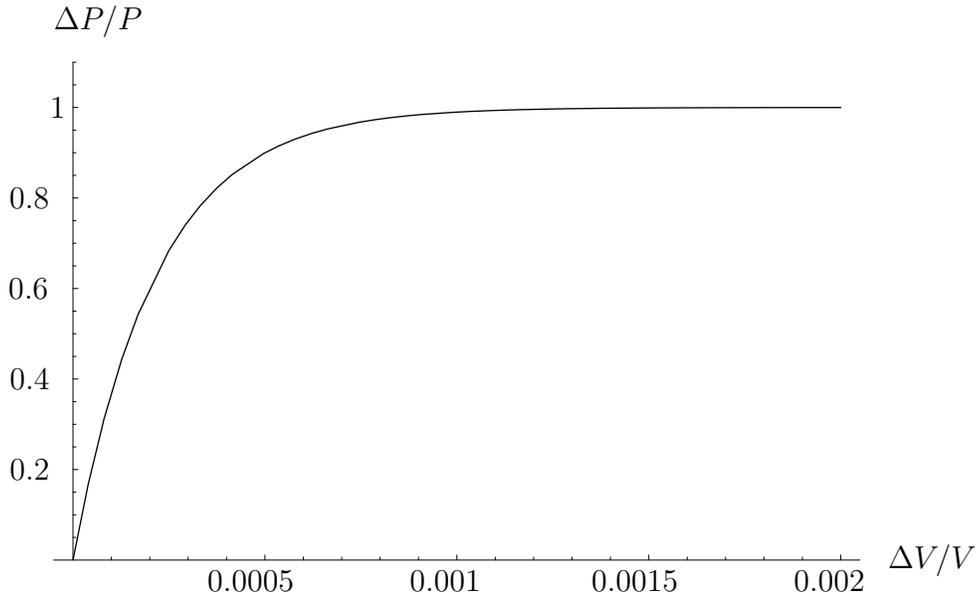,height=8cm} \caption{A plot of
probability, $\Delta P$, in a thin band of volume $\Delta V$ next to
the life-time boundary.  Here we considered the model $J=500$,
$q=10^{-2}$ and $\Lambda_0=1/2$.}
\label{boundfig}
\end{figure}

In order to establish the extent to which vacua near the life-time
boundary are favored we will consider a thin region next to it.  The
life-time boundary is at a constant value of $z$, and we will label
it by $z_1$.  The thin region we are considering next to this
boundary will extend down to a constant $z=z_2$, where $z_2<z_1$.
The volume contained within this region can be approximated by 
\be
\label{vol} \Delta V \simeq V_{J-2} \int_{z_2}^{z_1}
e^{-\frac{J}{2}z^2} dz, 
\ee 
where $V_{J-2}$ is the volume of a
$(J-2)$-sphere.  The probability of finding a vacuum in this region
can then be estimated using the formula (\ref{highlow}), and by
integrating over the thin band: 
\be \label{prob} \Delta P \simeq
\int_{z_2}^{z_1} \kappa_T d\Omega \simeq \int_{z_2}^{z_1}
e^{-\frac{J}{2}z^2-\frac{12}{z^2}} dz. \ee 
These integrals can be
performed numerically, the results of which are shown in figure
\ref{boundfig}.  Here we have normalized $\Delta P$ so that $\Delta
P/P \rightarrow 1$ as $z_2 \rightarrow 0$.  The volume $\Delta V$ is
presented as a fraction of the volume of the unit sphere.
\psfrag{DV}{$\Delta V/V$} \psfrag{DP}{$\Delta P/P$} \psfrag{1}{1}
\psfrag{0.8}{0.8} \psfrag{0.6}{0.6} \psfrag{0.4}{0.4}
\psfrag{0.2}{0.2} \psfrag{0.0005}{0.0005} \psfrag{0.001}{0.001}
\psfrag{0.0015}{0.0015} \psfrag{0.002}{0.002}
It can be seen from the figure that the probability is pressed up
against the boundary, within a fractional volume of $\sim 0.001$.

In order to determine the typical distance of an anthropically
acceptable vacuum from this boundary we will find the width of strip
that is required for half of the probability to be within it, and half
outside of it:  The half-life of the distribution.  The value of $z_2$
for such a band is $\sim 0.9995 z_1$.  The life-time of
vacua at this value of $z$ is $\sim 12.2$ Gyr.  Given the model we
have been investigating, with the parameters chosen and assumptions made, we should
therefore expect a typical anthropically acceptable vacuum to decay
after about $2.2$ billion years from the moment when the first observers appear
(at $10$Gyr).

One may have initially suspected that the exponential
pressure up against the boundary should cause decay of vacua after a
very short time.  However, the exponential sensitivity
of life-time on $z$ counters this pressure, and results in vacua being
reasonably long lived.

\section{Discussion}

In this paper we have considered the application of volume weighted
measures of eternal inflation to the Bousso-Polchinski landscape.  We
have found that the cosmological dynamics of volume-weighting results
in significant peaking of probability on the surface of the anthropic shell.
The anisotropy of tunneling rates in these models leads to the selection of vacua that
have one flux approximately ten times larger than the other 499 (for
$J=500$, $q=10^{-2}$ and $\Lambda_0=1/2$).  The peaking of
probability about this most likely outcome is very sharp, with
$99.9999\%$ of the probability on the anthropic shell expected to be in a region that is only
$\sim 10^{-23}$ of its surface.  The preferential selection of a
single large flux might suggest that  warped compactifications are
favored by this measure.  This is in contrast to the results of Bousso
and Yang \cite{boussoyang} who find that the holographic measure
\cite{bousso} selects fluxes along the diagonal of the flux space, with no one flux being especially
large.

Although this peaking gives us a sharp prediction for
the flux configuration we should expect from volume-weighted inflation, it still allows for a
satisfactory explanation of the cosmological constant problem.  The
reason for this is the very large number of points that are typically
in the anthropic shell, $\sim 10^{489}$ for the parameters chosen
above.  Even an area as small as $10^{-23}$ of the anthropic shell
still then contains $\sim 10^{466}$ vacua with anthropically acceptable values of the
cosmological constant\footnote{Other reasonable values of parameters
  also produce acceptably large numbers.}.  This is more than enough to produce the desired
smooth distribution in $\Lambda$ that is required for an
acceptable anthropic explanation of the problem.  We do not expect the
volume weighted measures of inflation to produce a staggered
distribution, such as that found in the bubble counting measure of \cite{bubbles}
for some values of parameters \cite{Schwartz-Perlov:2006hi}, \cite{Olum}.  

We have also found that the most likely vacua in the anthropic shell
often have a life-time too short for observers, such as ourselves, to
appear.  This additional anthropic pressure makes the vacua that are
sufficiently long-lived, and as close as possible to the favored
points, the most likely for observers to find themselves in.  Such a
configuration selects vacua that are long-lived enough for
observers to occur, but short-lived enough to be as close as possible
to the overall peak of probability.  We find that if it takes 10
billion years for observers to occur, then the most likely vacua for
them to find themselves in will decay a further 2.2 billion
years after this time, in our simple model.  This result is dependent
on the decay rates in section 2.2 being applicable to every vacuum in
the anthropic shell.  If anomalous vacua with longer life-times (due
to e.g. super-symmetry) are identified, then these vacua would be
immune to this selection pressure and may well prove to be more likely,
depending on their number and flux configuration.  

This apparent prediction of a doomsday might provide a plausible explanation of the
``coincidence problem'' of cosmology.  This problem, loosely stated,
asks why it should be that the energy densities observed in dark
energy and matter should be comparable at the present time.  If our
pocket of the universe is allowed to exist forever, then a randomly
distributed observer should expect to find themselves at a time when
the dark energy dominates, and all other energy densities are
negligible.  The existence of a doomsday, a few billion years after the
first observers appear, would change this picture and make it
considerably more likely for a random observer to find comparable
values of the dark energy and matter densities in their observable
universe.  A number of other solutions to this problem have also been
proposed in the literature.  For example, a compelling and model
independent solution exists in the context of holography \cite{Harnik}.

The results above have been calculated under the assumption that only
tunneling events to nearest neighbors in the landscape are possible.
However, it may be the case that multiple branes can be nucleated
simultaneously, in which case we would have to be re-evaluate our
results.  For example, according to the results of Garriga and Megevand
\cite{Garriga} the probability for simultaneously nucleating a stack
of $m$ branes, if possible, is approximately given by $\kappa^m$, where $\kappa$ is
the rate of nucleation of single branes.  If a result of this kind can
reliably be applied to the situation we have been considering here,
then it will have important consequences for our results -- such an
event, although unlikely, would be more likely than nucleating a
single brane, and then $m-1$ subsequent further branes from the lower energy vacua
produced.  To see this notice that the tunneling rate (\ref{highlow}), for transitioning from the highest
energy vacua to low energies, would become $\kappa_T \sim e^{-12/m z_0^2}$.
Here we have multiplied the numerator of the exponent by $m$, as
suggested by \cite{Garriga}, and multiplied the denominator by $m^2$, as the
larger jumps produced by nucleating multiple branes are equivalent to rescaling $z_0
\rightarrow m z_0$.  A large $m$ can now be seen to decrease the
exponent of $\kappa_T$, therefore reducing the peaking.  Clearly the possibility
of multiple brane nucleation events, if allowed to occur, will have a
potentially important effect.

Multiple brane nucleation events could also have a second important
effect, related to the minimum tunneling rate alluded to in section
2.2.  If it should be the case that any number of branes can be
nucleated simultaneously, then even the most unlikely events cannot
occur at a rate less than $e^{-8\pi^2/\Lambda}$, if they are allowed to
occur at all.  For vacua with
sufficiently large $\Lambda$ this rate is not negligible, and
tunneling events to distant corners of the landscape could occur with
a frequency comparable to that associated with vacua that are much closer.  Clearly, in
such a regime the sequences of tunneling events discussed above make
very little sense; instead one would expect a very nearly flat
distribution of probability over the entire landscape as the
highest energy vacua decay to all other vacua at approximately the same
minimum rate.  This is a very different picture to the one we have
been focusing up in this paper. If only one brane is
allowed to be nucleated at a time we have very strong peaking of
probability on the anthropic sphere, if, however, any number of branes
can be nucleated together then we have no peaking whatsoever.  As it
is this peaking that is responsible for a preference for short-lived
vacua, removing it also removes this pressure.  In
these situations it would be the combined effect of the many unlikely
transitions that would be of greatest importance.  

Another simplification that we have employed in this study is to
neglect the necessary period of slow-roll at the end of inflation, which is not
included explicitly in the Bousso-Polchinski landscape.  Linde argues in \cite{andreinew}
that volume weighted probabilities should be weighted by the amount of volume
produced during the slow roll period.  Such exponentially large factors could 
swamp the effects we have discussed here.  Clearly it will be important to have a
proper understanding of how to include slow-roll in this picture, in
order to gain a complete understanding of how
volume-weighted inflation proceeds in a more realistic landscape.

\vspace{10pt}

\leftline{\bf Acknowledgements}

We would like to thank R. Bousso, A. Linde, K. Olum, L. Susskind and
A. Vilenkin for very helpful comments and discussions.
This work is supported by NSF grant PHY-0244728.  TC acknowledges
the support of the Lindemann trust.

\appendix

\section{Changing the Location of the High-Energy Barrier}
\label{A1}

In section 4 the tunneling rates to the penultimate sphere, above
the anthropic one, were calculated in the limit of the high-energy
`smeared out' surface being at an infinite distance from the anthropic
surface.  From the results of section 3 it is clear that this
surface should, in fact, be at a finite distance, and so we must
justify our approximation of it being at infinity.  To achieve this
we will take some plausible values for the actual location of this
boundary, as suggested by section 3, and use them to evaluate the
quantities we are interested in.  We will find that the results of
our calculations are relatively insensitive to the precise location
of the high-energy surface, so that we are justified in
approximating it to be at infinity.

Let us first consider a boundary defined by the condition $3 \Delta
H = \kappa_{i \downarrow}$.  This condition can be considered as an
upper estimate for the high-energy surface:  Vacua that meet this
condition are still within the `smeared out' region.  Using the high
energy tunneling rate (\ref{r2}) for $\kappa_{\downarrow i}$, this
condition picks out a surface that can be used as an upper limit for
the integral (\ref{oops}).  This surface intersects the $z$-axis at
$z\simeq 1.23$, and drops to slightly lower $z$ as you move away
from the axis.

Alternatively, we may wish to consider a high-energy boundary
defined by $3 \Delta H = 10 \kappa_{i \downarrow}$.  Vacua that meet
this condition have violated the condition to be in the smeared out
region by an order of magnitude, and so can be considered to be
comfortably outside of the smeared regime.  This condition is then a
lower estimate for the high-energy surface.  Again, using (\ref{r2})
this condition picks out a surface which can be used as an upper
limit for the integral (\ref{oops}).  This lower estimate cuts the
anthropic shell, and intersects the $z$-axis at $z\simeq 0.61$.
Again, this surface drops to lower $z$ as you move away from the
axis.  The intersection of this surface with the anthropic shell
clearly gives a different picture - now tunneling to vacua in the
anthropic shell, above this surface, can be tunneled to in an
unsuppressed way.  Vacua below this surface must be reached in the
usual way, by a downward series of suppressed jumps.

Using these two estimates for the high-energy surface we can
numerically find the quantity we are most interested in:  The
saddle-point of the integral $\int \kappa_{\downarrow i} d \Omega$.
This quantity determines the position of the peak of probability,
and allows a straightforward estimate of the moments $\langle z^2
\rangle$ and $\langle x^2 \rangle$ (see section 4.3).  The location
of the saddle-point in these estimates, together with the case of a
high-energy boundary at infinity, are given in the table below.
\begin{table}[ht]
\begin{center}
\begin{tabular}{l|lll}
\hline\\[-7pt]
\shortstack[l]{Location of \\high-energy surface} & $\quad$
\shortstack{low\\($3\Delta H = 10 \kappa_{\downarrow i}$)} $\;$ & \shortstack{high\\($3\Delta H =
\kappa_{\downarrow i}$)} $\quad$ & \shortstack{infinity\\($z\rightarrow \infty$)} \\[5pt]
\hline\\[-7pt]
\shortstack[l]{Location of \\saddle-point} & $\qquad \; z_0=0.433$ & $\;\;\;
z_0=0.439$ & $z_0=0.440$ \\[5pt]
\hline
\end{tabular}
\end{center}
\end{table}

It can be seen that the location of the saddle-point is only weakly
dependent on the location of the high-energy surface.  The
difference between our higher estimate for the `smeared out'
surface, and taking this surface to be at infinity, is only at the
level of $\sim 0.2\%$. We attribute this to the fact that most of
the suppression is due to transitions at low energy, so including
extra transitions at high-energy gives only a modest contribution.
The difference between our lower estimate, and taking the
high-energy surface to infinity, is still relatively small at the
level of $\sim 1.6\%$.  This is surprisingly accurate considering
that in this approximation some of the transitions are completely
unsuppressed.  We attribute this accuracy to the unsuppressed points
near the $z$-axis contributing only a small amount to the total
probability flux, due to the small amount of area near the pole
compared to the large area away from it.  Further from the pole,
where the area is much larger, the high-energy surface is still far
enough away to be well approximated by being at infinity.

\section{Expectation Values Due to Tunneling From High Energy}
\label{appendixA}

To evaluate the expectation values $\langle z^2
\rangle$ and $\langle x^2 \rangle$, from equation (\ref{Aq}), consider the integral
\begin{align}
\label{Itoyq} I &= \int_0^\infty dz \prod_i^{J-1} \int_{-z}^z
dx_i (\sum_i x_i^2+z^2) e^{-(\sum_i x_i^2
  +z^2)} f^2 \kappa_T \\
&= \int_0^\infty r^{J+1} e^{-r^2} dr \int_\sigma f^2 \kappa_T
d\Omega
  \nonumber \\
&= \frac{1}{2} \Gamma\left(\frac{J+2}{2}\right) \langle
f^2\rangle \nonumber,
\end{align}
where $f=f(x_i,z)$ is any function of $x_i$ and $z$ on the surface of
the sphere.  In moving from the first to second line we have simply transformed
to a set of hyper-spherical polar coordinates, where $r \equiv
\sum_i x_i^2+z^2$.  In order to evaluate this second line, and
obtain the third, we have had to be careful in how we have defined
$\kappa_T$; specifically, we have required that $\kappa_T= \kappa_T
(\Omega)$.  This may at first seem contrary to (\ref{highlow}), but
it should be remembered that we are only interested in the value of
$\kappa_T$ on one particular surface -- that of the anthropic shell.  We
can then re-write (\ref{highlow}) as
\begin{equation} \nonumber
\kappa_T \sim e^{-12/\hat{z}^2},
\end{equation}
where $\hat{z}=\hat{z}(\Omega)$ is the value of $z$ on the unit sphere at the
point described by the polar coordinates $\Omega$.  For points not
contained on the unit sphere, $\kappa_T$ is obtained by connecting the
point to the origin with a radial vector and reading off the value
of $\hat{z}$ as this vector crosses the surface of the unit sphere.
Similar triangles then give
\begin{equation} \nonumber
\frac{\sum_i x_i^2+z^2}{1} = \frac{z^2}{\hat{z}^2}
\end{equation}
or
\begin{equation} \nonumber
\kappa_T \sim e^{-12\left(\sum_i
x_i^2+z^2\right)/z^2}.
\end{equation}
Similarly, we can re-write $x^2$ as
\begin{equation} \nonumber
\frac{\sum_i x_i^2+z^2}{1} =
\frac{x^2}{\hat{x}^2}
\end{equation}
where $\hat{x}=x(\Omega)$ is the corresponding value of $x$
at the point a suitable radial vector crosses the anthropic shell. With these
new expressions for $z^2$ and $x^2$, the integral (\ref{Itoyq}) can
be written as
\begin{equation} \nonumber
\langle \hat{f}^2\rangle=\frac{2}{\Gamma(\frac{J+2}{2})}
\int_0^\infty dz \prod_i^{J-1} \int_{-z}^z dx_i f^2 e^{-(\sum_i
x_i^2 +z^2)(1+12 z^{-2})}.
\end{equation}
Taking $f=z$ the integrals over the coordinates $x_i$ can all be
performed exactly, giving the moment $\langle z^2 \rangle$ as
\begin{equation}
\langle \hat{z}^2\rangle=\frac{2}{\Gamma(\frac{J+2}{2})} \int_0^\infty
dz z^2 \left[\frac{\sqrt{\pi}
  z}{\sqrt{12+z^2}} E(\sqrt{12+z^2}) \right]^{J-1} e^{-(12+z^2)}
\label{intq2}
\end{equation}
where $E(p)$ is the integral of a Gaussian distribution, or error
function, given by $2 \pi^{-1/2} \int_0^p e^{-q^2} dq$.  This
analysis has reduced $\langle z^2\rangle$ from a $J$ dimensional
integral, which is potentially very difficult to solve, to a one
dimensional integral that can be solved by saddle point
approximation, or by a simple numerical calculation.

We can proceed in a similar fashion to obtain an expression for
$\langle x^2\rangle$, which can also be expressed as a one-dimensional
integral. Such an analysis gives
\begin{align}\nonumber
\langle \hat{x}^2\rangle &=
\frac{2}{\Gamma(\frac{J+2}{2})} \int_0^\infty dz \prod_i^{J-1}
\int_{-z}^z dx_i x^2 e^{-(\sum_i x_i^2 +z^2)(1+12 z^{-2})}\\
\nonumber &= \frac{2}{\Gamma(\frac{J+2}{2})} \int_0^\infty dz
\int_{-z}^z dx x^2 \left[\frac{\sqrt{\pi}
  z}{\sqrt{12+z^2}} E(\sqrt{12+z^2}) \right]^{J-2}
e^{-(x^2+z^2)(1+12 z^{-2})}\\ \nonumber &=
\frac{2}{\Gamma(\frac{J+2}{2})} \int_0^\infty dz
 z^2 \left( \frac{1}{2 (12+z^2)}- \frac{e^{-(12+
     z^{2})}}{\sqrt{\pi} \sqrt{12+z^2} E(\sqrt{12+z^2})  } \right)
 \\ &\hspace{200pt} \times \left[\frac{\sqrt{\pi} z}{\sqrt{12+z^2}} E(\sqrt{12+z^2}) \right]^{J-1}
e^{-(12+ z^{2})}. \label{int2q2}
\end{align}
Going from the first to the second line, we have integrated over all
of the $x_i$ dimensions, except for $x$ (that is, the particular $x_i$
we are calculating the expectation value of).  Going from the second to
third lines we then integrate over $x$.

\section{The Single Peak Approximation for the Final Transition}
\label{appendixB}

The integral (\ref{A}) is not trivial to evaluate, and so we will
briefly consider a toy model in which the calculations are
significantly simpler to perform.  This model will allow us to fix
ideas and explicitly investigate approximations which will be of
direct use in solving (\ref{A}).  The toy model we propose is one in
which the rate of transitions goes as
\begin{equation}\label{gammatoy}
\kappa_{toy} \sim e^{-\hat{C}/z^2},
\end{equation}
where $\hat{C}$ is a constant.  This rate is clearly the same as
(\ref{highlow}), with a different numerical factor in the exponent.
We already know such models can be solved exactly, from the preceding
section.  We will use this knowledge to justify the single peak
approximation for the tunneling rate (\ref{tran}).

In order to show that the single peak approximation is valid for
(\ref{tran}), we will
proceed as follows.  First we will investigate this approximation in
our toy model.  Having shown the extent to which this
approximation is valid in our toy model, we will argue that it
should be a significantly better approximation in a more realistic
model.  We will then show the results of numerical calculations of
the more realistic model at low dimensions.  This will support our
previous argument and, when extended to larger dimensions, should
provide credible evidence that the single peak approximation is good.

\begin{figure}[t]
\center \epsfig{file=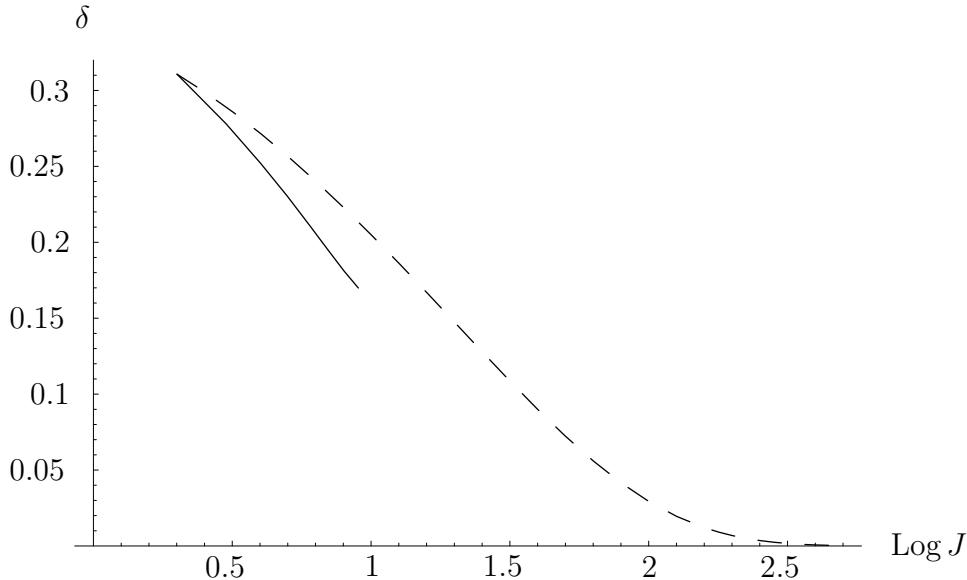,height=8cm} \caption{A plot of
$\delta$ as a function of dimensionality of the
  space, $J$, in the toy model (dashed line) and in a more realistic
  model (solid line).  The value of $\hat{C}$, in  the toy model is
  taken to be $0.7$, and the value of $\tilde{C}$, in the more
  realistic model is $0.3$.}
\label{deltatoy}
\end{figure}

For our toy model, the single peak approximation yields the
results as for (\ref{highlow}), with $200$ replaced by $\hat{C}$. We
now plot $\delta$ for the present case as a function of $J$ in
figure \ref{deltatoy}.
\psfrag{dd}{$\delta$} \psfrag{logJ}{$\textrm{Log}\,J$}
\psfrag{0.5}{0.5} \psfrag{1}{1} \psfrag{1.5}{1.5} \psfrag{2}{2}
\psfrag{2.5}{2.5} \psfrag{0.05}{0.05} \psfrag{0.1}{0.1}
\psfrag{0.15}{0.15} \psfrag{0.2}{0.2} \psfrag{0.25}{0.25}
\psfrag{0.3}{0.3}
It can be seen from the figure that the error, $\delta$, can be as
large as $\sim 0.3$ at low dimensionality ($J \rightarrow 2$). At
larger dimensionality, however, the value of $\delta$ can be seen to
drop off, and for $J = 500$ can be seen to be less than $0.03\%$.
We expect that this is due to the fact that at
higher dimensionality the boundaries of the domain of integration of
the actual region of interest extends to lower and lower values of
$z$.  This can be seen straight-forwardly from the expression for
the diagonal (that is, the place furthest from all axes where the
values of all coordinates are equal):
\begin{equation} \sum_i^J x_i^2
=1  \qquad \qquad \Rightarrow \qquad \qquad x_i \sim
\frac{1}{\sqrt{J}}.
\end{equation}
As this region extends to lower $z$, it covers more of the region
that is integrated over in the single peak approximation, making the
approximation increasingly more accurate as more dimensions are
added.

Having established that the single peak approximation in our toy model
is accurate to better than $0.03\%$, when $J = 500$,
we can begin to consider the accuracy of this approximation in more
realistic models.  We expect that the single peak approximation should
be more accurate in the more realistic models.  The reason for this
is that the rate of transitions is a steeper function of $z$
($e^{-1/z^3}$ instead of $e^{-1/z^2}$).  With this in mind we should
expect the extra area integrated over in the single peak approximation
to make even less of a contribution, as this extra area is at low
$z$.

The calculation to be performed in the more realistic model is
particularly cumbersome at large $J$, as we do not have the pleasant
reduction to a one-dimensional integral that was exhibited in the
toy model.  For low-dimensionality, however, numerical evaluations
can be performed.  The recipe in
this case is similar to that of the toy model.  Consider the
integral
\begin{align}
\label{I} I &= \int_0^\infty dz \prod_i^{J-1} \int_{-z}^z dx_i
(\sum_i x_i^2
  +z^2) f^2 e^{-(\sum_i x_i^2
  +z^2)^{3/2}} \kappa_F \\
&= \int_0^\infty r^{J+1} e^{-r^3} dr \int_\sigma f^2 \kappa_F d\Omega
  \nonumber \\
&= \frac{1}{3} \Gamma\left(\frac{J+2}{3}\right) \langle f^2\rangle
\nonumber.
\end{align}
Now, by the same method as before, we write the transition rate as a
function of angular coordinates
\begin{equation} \nonumber
\kappa_F \sim e^{-\tilde{C} \left(\sum_i x_i^2+z^2\right)^{3/2}/z^3}
\end{equation}
and
\begin{equation} \nonumber \frac{\sum_i x_i^2+z^2}{1} =
\frac{f^2}{\hat{f}^2}
\end{equation}
so that (\ref{I}) gives
\begin{equation} \label{A2}
\langle \hat{f}^2\rangle =\frac{3}{\Gamma(\frac{J+2}{3})} \int_0^\infty
dz
  \prod_i^{J-1} \int_{-z}^z dx_i f^2 e^{-(\sum_i x_i^2
  +z^2)^{3/2}(1+\tilde{C} z^{-3})}.
\end{equation}
In the single peak approximation this integral becomes
\begin{equation} \label{A3}
\langle\bar{f}^2\rangle =\frac{3}{\Gamma(\frac{J+2}{3})}
\int_0^\infty dz
  \prod_i^{J-1} \int_{-\infty}^\infty dx_i f^2 e^{-(\sum_i x_i^2
  +z^2)^{3/2}(1+\tilde{C} z^{-3})}.
\end{equation}
This last integral can be written in a significantly simpler way in
terms of a set of hyper-cylindrical coordinates (which has been the
reason for this whole subsection).  Here, however, this
simplification is not particularly beneficial as in order to
evaluate $\delta$ we must also analyze the more complicated integral
(\ref{A2}), over the actual region of interest.

The results of numerically integrating (\ref{A2}) and (\ref{A3}) are
also shown in figure \ref{deltatoy}.  Here $\delta$ is defined in
the same way as in the toy model, and we have set $\tilde{C}=0.3$.
The value of $\hat{C}$, for the toy model, was taken as $0.7$ in
order to match the two models at $J=2$ (the lowest dimensionality
for which $\delta$ is well defined).  It can be seen that the
magnitude of $\delta$ in this more realistic model is indeed smaller
than in the toy model, as conjectured earlier. Although we do not
have reliable estimates of $\delta$ at large $J$ in the more
realistic model (due to the difficulty of numerically integrating
over a large number of dimensions), we argue by analogy that we
should expect $\delta$ to continue dropping as $J$ is increased.

This concludes our argument in favor of the validity of the single
peak approximation.  We have shown that there exists a toy model in
which $\langle f^2\rangle$ can be evaluated in a more
straight-forward fashion. In this toy model we have shown that the
single peak approximation is accurate to $\sim 0.03 \%$ at $J =
500$, and that this accuracy increases as $J$ is increased.  We have
then argued that in our more realistic model the single peak
approximation should be more accurate than in the toy model.
Numerical investigations at low dimensionality indicate that this
expectation is correct, and comparison with the toy model leads us
to expect that the single peak approximation in more realistic
models should indeed be a good approximation at large $J$.


\begin{thebibliography}{99}

\bibitem{land}
  For a review see M.~Douglas and S.~Kachru,
  ``Flux Compactification,''
  [arXiv:hep-th/0610102].

\bibitem{Bousso:2000xa}
R.~Bousso and J.~Polchinski, ``Quantization of Four-Form Fluxes and
Dynamical Neutralisation of the Cosmological Constant,'' JHEP {\bf
0006}, 006 (2000) [arXiv:hep-th/0004134].

\bibitem{islands}  T. Clifton, A. Linde and N. Sivanandam, ``Islands
  in the Landscape,'' JHEP \textbf{0702}, 024 (2007) [arXiv:hep-th/0701083].

\bibitem{volumem}  A. Linde, ``Eternally Existing Self-Reproducing
  Chaotic Inflationary Universe,'' Phys. Lett. \textbf{B175}, 395 (1986).

\bibitem{andreinew}
A. Linde, ``Towards a Gauge Invariant Volume-Weighted Probability
Measure for Eternal Inflation,'' [arXiv:0705.1160].

\bibitem{bubbles}
J. Garriga, D. Schwartz-Perlov, A. Vilenkin and S. Winitzki,
``Probabilities in the Inflationary Multiverse,'' JCAP \textbf{0601},
017 (2006) [arXiv:hep-th/0509184].

\bibitem{Schwartz-Perlov:2006hi}
  D.~Schwartz-Perlov and A.~Vilenkin,
  ``Probabilities in the Bousso-Polchinski Multiverse,''
  JCAP {\bf 0606}, 010 (2006)
  [arXiv:hep-th/0601162].

\bibitem{bousso}
R. Bousso, ``Holographic Probabilities in Eternal Inflation,''
Phys. Rev. Lett. \textbf{97}, 191302 (2006) [arXiv:hep-th/0605263].

\bibitem{boussoyang}
R. Bousso and I.-S. Yang, ``Landscape Predictions from Cosmological
Vacuum Selection,'' [arXiv:hep-th/0703206].

\bibitem{vil}
J. Garriga and A. Vilenkin, ``Recycling Universe,'' Phys. Rev. D
\textbf{57}, 2230 (1998) [arXiv:astro-ph/9707292].

\bibitem{Linde:2006nw}
  A.~Linde,
  ``Sinks in the Landscape, Boltzmann Brains, and the Cosmological Constant
  Problem,''
  JCAP {\bf 0701}, 022 (2007)
  [arXiv:hep-th/0611043].

\bibitem{Coleman:1980aw}
  S.~Coleman and F.~De Luccia,
``Gravitational Effects on and of Vacuum Decay,''
  Phys.\ Rev.\ D {\bf 21}, 3305 (1980).

\bibitem{Park}
S. Parke, ``Gravity and the Decay of False Vacuum,''
Phys. Lett. \textbf{121B}, 313 (1983).

\bibitem{brown}
J. Brown and C. Teitelboim, ``Dynamical Neutralization of the
Cosmological Constant,'' Phys. Lett \textbf{B195}, 177 (1987).

\bibitem{Garriga}
J. Garriga and A. Megevand,
``Coincident Brane Nucleation and Neutralisation of $\Lambda$,''
Phys. Rev. D \textbf{69}, 083510 (2004)
[arXiv:hep-th/0310211].

\bibitem{kklt}
S. Kachru, R. Kallosh, A. Linde and S. Trivedi,
``De Sitter Vacua in String Theory,''
Phys. Rev. D \textbf{68}, 046005 (2003)
[arXiv:hep-th/0301240].


\bibitem{ccbound}
A.~Linde, ``The Inflationary Universe,'' Rept.\ Prog.\ Phys.\  {\bf 47},  925 (1984),
T.~Banks,
  ``T C P, Quantum Gravity, the Cosmological Constant and
All That..,''
  Nucl.\ Phys.\  B {\bf 249}, 332 (1985),
S. Weinberg, ``Anthropic Bound on the Cosmological Constant,''
Phys. Rev. Lett. \textbf{59}, 2607 (1987). 

\bibitem{ads}
A. Linde, ``Inflation and Quantum Cosmology,'' Print-86-0888 (June
1986), in {\it 300 Years of Gravitation}, ed. by S. Hawking and
W. Israel, Cambridge University Press, Cambridge (1987).

\bibitem{Olum}
K. Olum and D. Schwartz-Perlov, ``Anthropic Prediction in a Large Toy
Landscape,'' [arXiv:0702.2562].

\bibitem{Harnik}
R. Bousso, R. Harnik, G. Kribs and G. Perez, ``Predicting the
Cosmological Constant from the Causal Entropic Principle,'' [arXiv:hep-th/0702115]. 

\end{thebibliography}
\end{document}